\documentclass[structabstract]{aa}
\usepackage{natbib}
\bibpunct{(}{)}{;}{a}{}{,}

\usepackage{graphicx}
\usepackage{lscape}
\usepackage{aalongtable}
\usepackage{subfigure}

\usepackage{txfonts}
\usepackage{comment}
\usepackage{color}
\usepackage{ifthen}

\newcommand{\Ha}{\mbox{H$\alpha$}}
\newcommand{\Hb}{\mbox{H$\beta$}}

\specialcomment{nccomments}{\begingroup\sffamily\color{blue}}{\endgroup}

\begin{document}

\title{Understanding star formation and feedback in small galaxies\thanks{Based on observations made with ESO Telescopes at Paranal Observatory under program ID~079.B-0445.}}

\subtitle{The case of the blue compact dwarf Mrk~900}

\author{L. M. Cair\'os
    \inst{1}
    \and
    J.N. Gonz\'alez-P\'erez
    \inst{2}}
\institute{Institut f{\"u}r Astrophysik, Georg-August-Universit{\"a}t,
Friedrich-Hund-Platz 1, D-37077 G{\"o}ttingen, Germany \\
           \email{luzma@astro.physik.uni-goettingen.de}
           \and Hamburger Sternwarte,
            Gojenbergsweg 112,
21029 Hamburg\\
\email{jgonzalezperez@hs.uni-hamburg.de}
}

\date{Received ...; accepted ...}


\abstract
{Low-luminosity, active star-forming blue compact galaxies (BCGs)
are excellent laboratories for  investigating the process of  
star formation on galactic scales and to probe the interplay between 
massive stars and the surrounding interstellar (or intergalactic) medium.}
{We  investigate the morphology, structure, and stellar content of   BCG Mrk~900, as well as the 
excitation, ionization conditions, and kinematics of its  \ion{H}{ii} regions and surrounding ionized gas.
}
{We obtained integral field observations of Mrk~900 working with the Visible
Multi-Object Spectrograph at the Very Large Telescope. The observations were taken 
in 
the wavelength range 
4150-7400\,\AA\, covering  
a field of view of
27$\arcsec \times$ 27$\arcsec$ on the sky with a spatial sampling of 
$0\farcs$67. From the integral field data we  built continuum, emission, and 
diagnostic line ratio maps and produced  
velocity and velocity dispersion maps.  We also generated  the integrated spectrum of
 the major
\ion{H}{ii} regions and the nuclear area to determine reliable physical
parameters and oxygen abundances. Integral field spectroscopy was complemented with
deep broad-band photometry taken at the 2.5~m NOT telescope; the broad-band data, tracing  the galaxy up to radius 4~kpc, allowed us to 
investigate the properties of the low surface brightness underlying stellar host.}
{We disentangle two different stellar components in Mrk~900: a young population, which resolves into individual stellar clusters with ages $\sim$5.5-6.6~Myr and extends about 1~kpc along the galaxy minor axis,  is placed on top of a   
rather red and regular shaped underlying stellar 
host, several Gyr old.
We find evidence of  a substantial amount of dust and an inhomogeneous extinction
pattern, with a dust lane crossing the central starburst. Mrk~900 displays overall rotation, although distorted in the central, starburst regions;  the dispersion velocity map  is highly inhomogeneous, with values increasing up to 60~km~s$^{-1}$ at the
periphery of the SF regions,  where we also find hints of the presence of shocks. Our observational results point  to an interaction or merger  with a low-mass object or infalling gas as plausible trigger mechanisms for the present starburst event. 
}


\maketitle
%

\section{Introduction}

Blue compact galaxies (BCGs) are low-luminosity (M$_{B}\geq$-18) 
and small systems (starburst radius $\leq$~1~kpc) that form stars at unusually high rates (up to 3~M$_\odot$~yr$^{-1}$; \citealp{ThuanMartin1981,Cairos2001a,Cairos2001b, Fanelli1988,HunterElmegreen2004}).  They are also rich in gas
(M$_{HI}$=10$^{8}$-10$^{9}$M$_\odot$) and present  low
metal abundances, as derived from their warm ionized gas 
(1/40~Z$_\odot\leq$~Z~$\leq$1/2~Z$_\odot$; \citealp{Thuan1999,Salzer2002,Izotov1999,Kunth2000}). These
characteristics make them excellent targets  for  investigating the process of
star formation (SF) in galaxies:

\smallskip

 First, the lack of spiral density waves and strong
shear forces (the mechanisms assumed to trigger and maintain SF in spirals;
\citealp{Shu1972,Nelson1977,Seigar2002}) allow us to investigate the SF process in a
relatively simple environment and to search for alternative trigger mechanisms
\citep{Hunter1997}; for instance, there is evidence that feedback from massive
stars  could be responsible for the ongoing starburst in several BCGs 
\citep{Cairos2017a,Cairos2017b}.  

\smallskip

Second, the impact of massive stars into the  interstellar medium (ISM) of a low-mass
galaxy can be dramatic: as the blast-waves created by supernova (SN) explosions
propagate, they give rise to huge expanding shells \citep{McCray1987}, which in the
absence of density waves and shear forces grow to larger sizes and live longer than in
typical spirals. Observations of dwarf star-forming galaxies reveal ionized gas structures stretching
up to kiloparcec scales \citep{Hunter1990,Marlowe1995,Bomans1997,Bomans2007,Martin1998,Cairos2001b,Cairos2015}. In a shallow 
potential well  these expanding shells can break out of the galaxy disk, or even the
halo, and can vent the SN enriched material into the ISM and/or intergalactic medium
(IGM; \citealp{Dekel1986,Marlowe1995,Martin1998,Martin1999,MacLow1999}).

\smallskip 

Finally, the
low metal content and high star formation rate (SFR) of BCGs give us the opportunity to
characterize  
starburst events that 
are taking place in conditions very similar to those of the
early Universe  \citep{Madden2006,Madden2013,Lebouteiller2017}.

\smallskip 

Motivated by the relevance of these topics 
we initiated a project focused on
BCGs and, in particular, on their current starburst   
episode and the impact on the ISM. 
To this end we took integral
field spectroscopy (IFS) of a sample of forty galaxies. 
The first results from our analysis are presented in 
\cite{Cairos2009a,Cairos2009b,Cairos2010,Cairos2012,Cairos2015}. 
A complete understanding of the SF process in low-mass systems and  on the complex interaction of massive stars with their environment demand also detailed analyses of individual 
objects \citep{Cairos2017a,Cairos2017b}.
Here we focus on the BCG Mrk~900 and 
combine IFS observations with optical imaging
to investigate its recent SF and evolutionary history.

\smallskip

Mrk~900 (NGC~7077) is a
relatively luminous BCG (M$_{B}$=-17.08; this work), included in the \cite{Mazzarella1986}
catalog of Markarian galaxies. Surface photometry in the optical and
near-infrared (NIR) revealed a blue starburst on top of a redder regular host 
\citep{Doublier1997,Doublier1999,GildePaz2003,GildePaz2005,Micheva2013a,Janowiecki2014}.
\cite{Cairos2015} investigated
the starburst component of Mrk~900 by means of IFS: 
the galaxy
emission-line maps showed that SF occurs   in various knots, aligned on a
southeast--northwest axis,  with the largest \ion{H}{ii} region displaced 
northwest. The distinct morphology in continuum maps, which peaks at the galaxy
center are suggestive of different episodes of SF. In addition, 
holes and filaments in the ionized gas, together with strong low-ionization
lines ([\ion{O}{i}]~$\lambda6300$ and
[\ion{S}{ii}]~$\lambda\lambda6717,\,6731$) revealed an important impact of 
the SF 
 on the surrounding ISM and suggest the presence of shocks. For all these 
 reasons Mrk~900 appears to be an ideal target for  a thorough study of
the effect of SF on the ISM of a dwarf galaxy.

\begin{table}
\caption{Basic parameters of Mrk~900}
\begin{center}
\begin{tabular}{lcc}
\hline
Parameter   & Data & Reference \\ 
\hline
\\
Other names        & NGC~7077, UGC~11755  &  \\
 RA (J2000)        &  21$^h$29$^m$59$\fs$6  &      \\
 DEC (J2000)       &  02$\degr$24$\arcmin$51$\arcsec$     &      \\
 V$_{hel}$         &  1152$\pm$5 km~s$^{-1}$ & \\
 Distance          &  18.9$\pm$1.3      Mpc  &     \\
 Scale             &   91 pc~arcsec$^{-1}$           &    \\                     
 D$_{25}$          &    49.9$\pm$0.06 $\arcsec$ (4.54~kpc)  &    RC3   \\
 A$_{B}$           &       0.211                   &  SF11  \\
 M$_{B}$           & -17.07           & This work \\
 M$_{*}$            & 9.5$\times$10$^{8}$M$\odot$& H17  \\
 M$_{HI}$          & 1.55$\times$10$^{8}$M$\odot$& VSS01        \\
  M$_{DYN}$      & 1.64$\times$10$^{9}$M$\odot$& VSS01  \\
D$_{HI}$/D$_{25}$ & 1.2 & VSS01\\ 
  Morphology        & SO$^{-}$pec?; BCD    & RC3; GP03\\
 \hline
\end{tabular}
\end{center}
Notes:
RA, DEC, heliocentric velocity, distance, scale, and Galactic extinction are  from
NED (http://nedwww.ipac.caltech.edu/). The distance was calculated using a Hubble constant of 73 km s$^{-1}$ Mpc$^{-1}$,
and taking into account the influence of the Virgo Cluster, the Great
Attractor, and the Shapley supercluster.
HI-to-optical diameter (D$_{HI}$/D$_{25}$) is measured at the 10$^{20}$ atoms~cm$^{-2}$ and 
25~mag~arcsec$^{-2}$ isophotes. 
References:   GP03 =
\cite{GildePaz2003};   H17=\cite{Hunt2017}; RC3= \cite{deVaucouleurs1991}; SF11=\cite{Schlafly2011}; VSS01=\cite{VanZee2001}.
\end{table}

\section{Observations and data processing}

We  carried out a spectrophotometric analysis of the BCG Mrk~900. 
The galaxy central starburst was studied  by
means of IFS, whereas 
deep broad-band imaging in the optical was  used to derive the properties of the 
underlying host galaxy.  Details of the observations are provided Table~\ref{Table:log}.

\begin{table}
\caption{Log of the observations}
\begin{center}
\begin{tabular}{cccccc}
\hline
Date         & Tel. & Inst.     & Grism/Filter & Exp.   & Seeing\\
             &      &           &              &     (s)     &  ($\arcsec$)  \\    \hline
Aug. 2005   & NOT             & ALFOSC      & B            & 2700               & 0.8\\
Aug. 2005   & NOT             & ALFOSC      & V            & 2700               & 0.6\\
Aug. 2005   & NOT             & ALFOSC      & R            & 2700               & 0.7\\
Aug. 2007   & VLT             & VIMOS       & HR-Blue      & 4320               & 0.9-1.4\\
Aug. 2007   & VLT             & VIMOS       & HR-Orange    & 4320               & 0.5-0.9\\
\hline
\end{tabular}
\end{center}
Notes: The columns list, respectively, observation date, telescope and instrument used, grism or filter,
total exposure time and seeing.
\label{Table:log}
\end{table}

\subsection{Broad-band imaging}

\begin{figure*}[h]
\centering
\includegraphics[angle=0, width=1.0\linewidth]{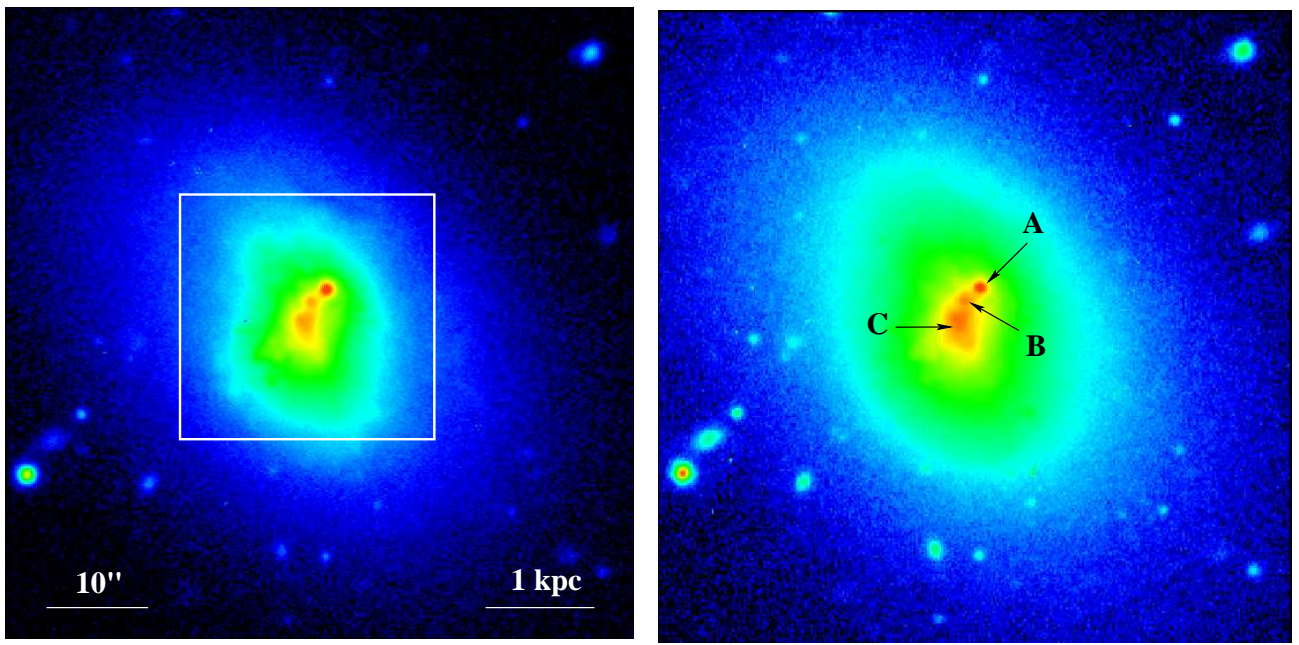}
\caption{Broad band images of Mrk~900. Left:  B-filter image; the FoV is  
$\sim1.0\arcmin\times1.0\arcmin$ ($5.5\times5.5$~kpc$^{2}$) and  
the central box
indicates the $27\arcsec\times27\arcsec$ ($2.5\times2.5$~kpc$^{2}$) VIMOS FoV. 
 Right: R-band image (same FoV), with the  central major
stellar clusters marked. 
North is up and east to the left.}
\label{Figure:mrk900_not} 
\end{figure*}

Broad-band observations of Mrk~900 were carried out  in August 2005 
with  the Nordic Optical Telescope (NOT) at Observatorio del Roque de los
Muchachos (ORM) in La Palma. The telescope was equipped with ALFOSC (Andaluc\'ia
Faint Object Spectrograph and Camera), which provided a field of view (FoV) of
6.5$'\times$6.5$'$ and a scale of 0.188$"$~pix$^{-1}$.
We collected CCD images through the
$B$, $V$,  and $R$ filters. We observed under excellent weather conditions: the 
seeing varied 
between 0.6 and 0.8~arcsec and the night was photometric.

\smallskip 

Image processing was carried out using standard {\sc iraf}\footnote{The Image Reduction and Analysis Facility (IRAF) is a software system
for the reduction and analysis of astronomical data. It is distributed by the NOAO, which is
operated by the Association of Universities for Research in Astronomy, Inc., under cooperative
agreement with the National Science Foundation} procedures. Each
image was corrected for  bias, using an average bias frame, and was flattened
by dividing by a mean  twilight flat-field image. The average sky level was
estimated by computing the mean value within several boxes surrounding the
object, and subtracted out as a constant. The frames were then registered (we took a set of three dithered exposures for
each filter) and combined to obtain
the final frame,  with cosmic ray events removed and bad pixels cleaned out. 
Flux calibration was performed through the observation of photometric stars from  
the \citet{Landolt1992} list.

\subsection{Integral field spectroscopy}

Spectrophotometric data of Mrk~900 were collected at the Very Large Telescope
(VLT; ESO Paranal Observatory, Chile), working with the \emph{Visible
Multi-Object Spectrograph} (VIMOS; \citealp{LeFevre2003}) in its integral field
unit (IFU) mode. The observations were done in Visitor Mode in August 2007 with
the blue (HR-Blue) and orange (HR-Orange) grisms in high-resolution mode
(dispersion of 0.51\,\AA\,pix$^{-1}$ in the wavelength range of
4150--6200\,\AA, and dispersion of 0.60\,\AA\,pix$^{-1}$  in the range
5250--7400\,\AA). A field of view (FoV) of 27$\arcsec \times$ 27$\arcsec$ on
the sky was mapped with a spatial sampling of  $0\farcs$67. The weather
conditions were good (see  Table~\ref{Table:log}), and galaxy exposures were
taken at airmass 1.12-1.21. The spectrophotometric standard EG~274 was observed
for flux calibration.

\smallskip 

Data were processed using the ESO VIMOS pipeline (version 2.1.11) via the graphical user
interface {\sc Gasgano}. The observations and a complete description of the data reduction
are presented in \cite{Cairos2015}.

\subsubsection{Emission-line fitting}
\label{linefitting}

The next step in the IFS data process is the measurement of the fluxes in
emission lines,  required to produce the  bidimensional galaxy maps. We computed
the fluxes by fitting Gaussians to the line profiles.  The fit was
performed using the task {\tt fit} of {\sc Matlab} with the 
{\em Trust-region} algorithm for nonlinear least squares.
 We ran an automatic procedure, which fits a series of lines for every spaxel, 
namely, H$\beta$, [\ion{O}{iii}]~$\lambda4959$, [\ion{O}{iii}]~$\lambda5007$, [\ion{O}{i}]~$\lambda6300$, 
H$\alpha$, [\ion{N}{ii}]~$\lambda6584$ and
[\ion{S}{ii}]~$\lambda\lambda6717,\,6731$. For each line the fitting 
task provides the 
flux, the centroid position, the line width and continuum, and the corresponding
uncertainties of each parameter; 1~$\sigma$ errors are computed by the task
using the inverse factor 
from QR decomposition of the Jacobian, the degrees of freedom, 
and the root mean square. After carefully inspecting the data, in particular in regions with 
constant emission-line fluxes, we found that the errors given by the fitting task seem underestimated. Therefore, we increased the errors of the emission-line fluxes 
by multiplying by 1.34, a factor that matches the spatial variations of the fluxes.  

\smallskip

Line profiles were fitted using a single Gaussian in
all but the  Balmer lines.  Determining accurate fluxes of the Balmer emission
lines is not straightforward since these fluxes can be significantly
affected by stellar absorption 
\citep{McCall1985,Olofsson1995,GonzalezDelgado1999b}.  To take this effect into
account,  we   applied two different methods depending on
the characteristics of the individual profile. If the
absorption wings were clearly visible (as was often the case in  H$\beta$), we
used two Gaussians, one in emission and one in absorption, and 
derived the fluxes in absorption and emission simultaneously. In the absence
of clear visible absorption wings, a reliable decomposition is impossible (this is 
often the case with H$\alpha$), and we assumed
that the equivalent width in absorption in H$\alpha$ was the same as in
H$\beta$; this is well supported by the predictions of the models
\citep{Olofsson1995,GonzalezDelgado1999b}.

\smallskip 

In low surface brightness (LSB) regions a spatial smoothing procedure was
applied to increase the accuracy of the fit. Depending on the signal-to-noise
ratio (S/N), the closest 5, 9, or 13 spaxels were averaged before
the fit was carried out. In this way, we maintain the spatial resolution of the
bright regions of the galaxy while obtaining a reasonable S/N for the faint
parts, but with a lower spatial resolution.

\subsubsection{Creating the galaxy maps}
\label{creatingmaps}

The  parameters of the line fits were used subsequently to construct the 2D maps,
taking advantage of the fact that the combined VIMOS data are arranged in a
regular 44$\times$44 matrix.  

\smallskip

Continuum maps in different spectral ranges were obtained by integrating the flux
in specific spectral windows, selected so as to avoid strong emission
lines or residuals from the sky spectrum subtraction. We also built an integrated
continuum-map  by integrating over the whole spectral range, but
masking the spectral regions with a significant contribution of emission lines.

\smallskip

Line ratios maps for lines falling in the wavelength range of either grism were
computed by dividing the corresponding flux maps. In the case of  
H$\alpha $/H$\beta$,  the line ratio map was derived after registering and
shifting the H$\alpha$ map to spatially match the H$\beta$ map. The 
shift was calculated using the difference in position of the center of the
brighter \ion{H}{ii} regions.  The shift was applied using a bilinear
interpolation. In order to correct for the fact that  H$\alpha$ and H$\beta$
had been observed under different seeing conditions,  the H$\alpha$ map was degraded convolving with a Gaussian to match both PSFs. 

\smallskip 

Only spaxels with a flux level higher than the
3$\sigma$ level  were considered when building the final maps. All maps were corrected for interstellar
extinction in terms of spaxels applying Eq.~\ref{ext_eq} (see
Sect.~\ref{extinction}).

\section{Results}
\subsection{Broad-band morphology: a first view on the stars}
\label{morphologyandstars}

\begin{figure}[h]
\centering
\includegraphics[angle=0, width=0.9\linewidth]{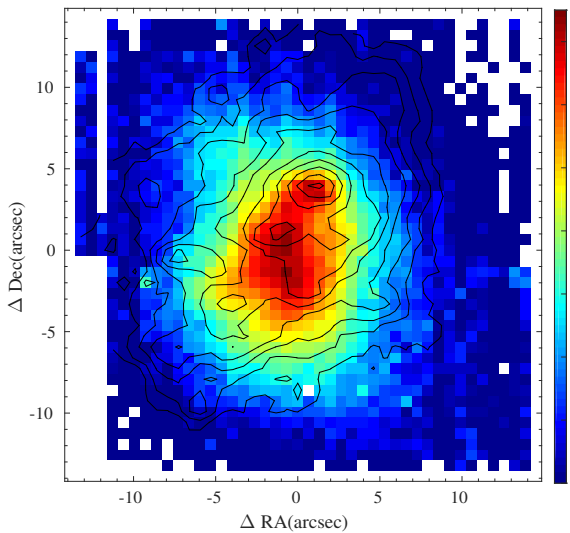}
\caption{Mrk~900 continuum map from the VIMOS/IFU data built by summing over the whole blue spectral range  (4150--6200\,\AA), but masking the emission lines. Contours in H$\alpha$ are overplotted. The 
scale is logarithmic and the units are arbitrary.  The FoV is $\sim$2.5$\times$2.5~kpc$^{2}$, with a spatial
resolution of about 61~pc per spaxel. North is up and east to the left; the axes represent the  displacements in RA and DEC with respect to the center of the FoV, also in all the maps presented from here on.}
\label{Figure:mrk900_cont} 
\end{figure}

The NOT B and R frames of Mrk~900 are presented in
Figure~\ref{Figure:mrk900_not}. In the full ALFOSC FoV the irregular high
surface brightness (HSB) region is very well distinguished on top of  the low surface
brightness (LSB) host.  The HSB area is resolved into three major knots ({\sc
a}, {\sc b}, and {\sc c} in Figure~\ref{Figure:mrk900_not}); the emission peaks
at the position of the knot~{\sc a} (the clump displaced northwest), while
knot~{\sc c} is situated roughly at the center of the redder host. The VIMOS continuum
map (Fig.~\ref{Figure:mrk900_cont}) essentially reproduces  the same morphological
pattern of the galaxy in broad-band frames, but with a notable difference: in
continuum the intensity peaks at the position of knot~{\sc c}, and knot~{\sc
a} is just a moderate emitter. We  masked strong emission lines when constructing
the VIMOS continuum map; however, broad-band filters also encompass  the emission from
the ionized gas\footnote{The B-band filter is mostly affected  by the
H$\gamma$ and  H$\beta$ lines, whereas the emission from 
H$\alpha$, [SII], and [NII]
fall in the R band.}. The much higher intensity of knot~{\sc a} in
broad-band filters  compared with the continuum indicates that a significant
part of its light  originates  in the ionized gas and reveals the
presence of ionizing (very young) stars.

\smallskip

We built the galaxy color maps (i.e., the
ratio of the flux emitted in two different filters). We computed the maps from the flux calibrated 
broad-band images after they were aligned and matched to the same seeing. In order to  improve the
S/N in the outer galaxy regions, we applied a circular-averaging filter to the individual images,
the radius of the filter depending on  the flux level in the R band; the filtering is only
implemented at large radius ($\geq 10"$), so  that the spatial resolution in the central
HSB regions is preserved. To avoid artifacts we did not remove foreground stars
from the field.

\smallskip

\begin{figure*}
\centering
\includegraphics[angle=0, width=0.9\linewidth]{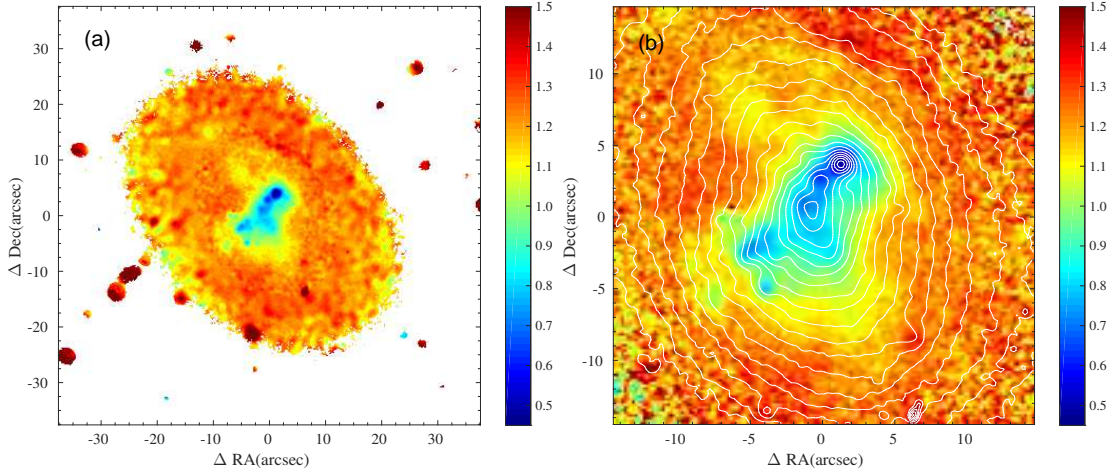}
\caption{(a) $(B-R)$ color map of Mrk~900 derived from the NOT frames.
(b) Zoomed-in image of the central 27$"\times$27$"$ (VIMOS FoV) of the $(B-R)$ color map, with the R-band contours overlaid.}
\label{Figure:mrk900_br} 
\end{figure*}

Two 
major stellar components are clearly distinguished in the $(B-R)$ color map of
Mrk~900 (Figure~\ref{Figure:mrk900_br}).  The 
HSB region appears distorted and much bluer (most probably  the result of one  or
more episodes of SF), while the host galaxy shows a smooth elliptical
shape with a markedly redder color. There is an apparent color gradient among the major
central knots: $(B-R)\sim$0.23, 0.56, and 0.64 in knots~{\sc a}, {\sc b}, and ~{\sc c}, respectively, corrected from Galactic extinction
following \cite{Schlafly2011}.  We find for the 
galaxy host a roughly constant  
$(B-R)\sim$1.21, in good agreement with the
color profiles reported by \cite{GildePaz2003} and
\cite{Micheva2013a}.

\subsection{Integrated and surface brightness photometry} 
\label{photometry}

The photometry of Mrk~900 is presented in Table~\ref{tab:photometry}. 
We computed the integrated magnitudes as the magnitudes within the isophote at
surface brightness 26.5~mag arcsec$^{-2}$ in the B band, namely 
the Holmberg isophote. We
corrected for Galactic extinction using the reddening coefficients from 
\cite{Schlafly2011}.

\begin{table}
\caption{Integrated photometry and structural parameters of Mrk~900.\label{tab:photometry}}
\begin{center}
\begin{tabular}{lccc}
\hline
Parameter   & B & V  & R \\ 
\hline
m (mag)                                                         &  14.30$\pm$0.03         & 13.68$\pm$0.04                  & 13.35$\pm$0.03       \\
M (mag)                                                          & -17.08                          & -17.70                               & -18.03  \\
\hline
$\mu_{0,host}$~(mag~arcsec$^{-2}$)         & 21.60$\pm$0.10     &   20.87$\pm$0.10                 & 20.20$\pm$0.06   \\
$\alpha_{host}$~(pc)                             & 726$\pm$25                   &  754$\pm$24                     &  673$\pm$13\\
m$_{host}$ (mag)                                         &  14.89                       & 14.12                                  & 13.73      \\
M$_{host}$ (mag)                                                 &  -16.49                         & -17.26                                & -17.65    \\
\hline
\end{tabular}
\end{center}
Notes. Integrated magnitudes computed from the flux within the isophote at 26.5~mag per 
arcsec$^{-2}$.  Structural parameters of the exponential disk model best fitting  the underlying host galaxy and integrated and absolute magnitudes of the host;   
all magnitudes are corrected from Galactic extinction
following  \cite{Schlafly2011}. 
\end{table}

\smallskip 

 We also built the galaxy surface brightness profiles (SBPs). There are many studies 
on surface brightness photometry of BCGs 
\citep{Papaderos1996,Papaderos2002,Doublier1997,Doublier1999,Cairos2001b,Noeske2003,Noeske2005,GildePaz2005,Caon2005,Micheva2013a,
Micheva2013b,Janowiecki2014}. As  pointed out in most of them, the construction of the SBP
of a BCG is not
straightforward;  the most common methods assume some sort of  galaxy symmetry and do not  necessarily apply to the irregular HSB region of a BCG 
\citep{Papaderos1996,Papaderos2002,Doublier1997,Cairos2001a,Noeske2003,Noeske2005}.  A simple way
to overcome this problem is to construct the profile using different approaches in the high-
and low-intensity regimes; in the brighter regions, a method that does not  require any assumption on
the galaxy morphology is used, whereas ellipses are fitted to the isophotes of the smooth host
\citep{Cairos2001a}.

\smallskip

Here we opted to build the light profiles 
using standard techniques,  but bearing in mind
that they provide a 
proper description of the galaxy structure 
only outside the starburst region, which is easily assessed using the color map (Figure~\ref{Figure:mrk900_br}). After masking out 
foreground
stars and nearby objects, we fitted elliptical isophotes to the images  using
the IRAF task {\tt ellipse}  \citep{Jedrzejewski1987}. The fit was done 
in two steps: first, the parameters of the ellipses were left free to vary;  in a 
second run, and  to improve the stability of the final fit,  we fixed the center
to the average center of the outer isophotes found in the first step. We found 
that at large radius (20"$\leq$~r~$\leq$40") the galaxy isophotes are well fitted with ellipses of
position angle PA=40$^{\circ}$ and ellipticity $\epsilon$=0.30, in
good agreement with the values PA$=41^{\circ}$ and  $\epsilon=$0.27  
reported by  \cite{Micheva2013a} . 

\smallskip

\begin{figure}
\centering
\includegraphics[angle=0, width=0.8\linewidth]{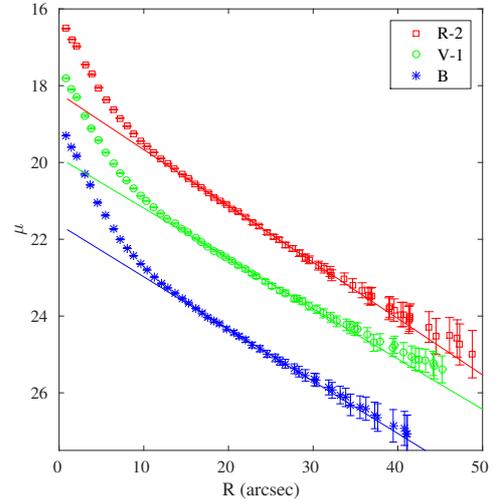}
\caption{Surface brightness profiles of Mrk~900 in B, V, and R. For a better visualization the V and R profiles are shifted 1 and 2 magnitudes, respectively. 
We also plotted the straight lines describing the exponential profile of the underlying host galaxy.}
\label{Figure:SBP}
\end{figure}

The SBPs of Mrk~900 in $B$, $V$, and $R$ are displayed in
Figure~\ref{Figure:SBP}. The error bars  on the surface brightness magnitudes
were calculated taking into account the errors given by the fit and the 
errors on the estimation of the sky value.  The
profiles 
show the common behavior among BCGs:  there is a  brightness excess
in the inner region,  but 
at  large radii  the
intensity  seems well described by an exponential decay.

\smallskip

We fitted an exponential
disk to the outer profile to obtain the structural parameters of the underlying host galaxy. We performed a weighted linear
least-squares fit to the function

\begin{equation}
\mu(r)=\mu_{0}+1.086\frac{r}{\alpha}
\label{mu_eq}
,\end{equation}

\noindent where $\mu$(r) is the observed surface brightness at radius r, $\mu_{0}$ is the extrapolated 
central surface brightness, and $\alpha$ is the exponential scale of the disk. 
A delicate step is the selection of the fitting radial range: the 
lower radial  limit is given by the position where the contribution of the starburst light is
negligible, and the higher radius is set by the point where the sky errors are very large or the ellipses parameters  unstable; the final values of the
parameters are not very sensitive to the higher radius used since the larger
errors make the weight of these points to be very low.  The structural parameters
resulting from the  fit in each band (i.e., the central surface brightness, $\mu_{0,host}$, and the scale
length, $\alpha_{host}$) are  presented in Table~\ref{tab:photometry}.
Adopting the disk model, we also computed 
the integrated and absolute magnitudes of the host galaxy.

\smallskip

\subsection{Emission-line maps:  warm ionized gas}
\label{vimoslinemaps}

\begin{figure}
\centering
\includegraphics[angle=0, width=0.9\linewidth]{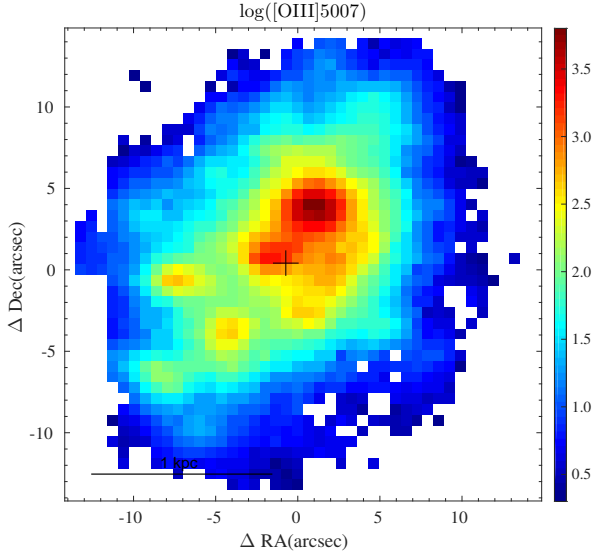}
\caption{[\ion{O}{iii}]~$\lambda5007$ emission-line flux 
map for Mrk~900, with a cross marking the position of the 
continuum peak. The spatial scale in pc is indicated at the 
bottom left; the spatial resolution is about
61~pc per spaxel (91~pc per arcsec) and flux units are $10^{-18}$\,erg\,s$^{-1}$\,cm$^{-2}$, also in all the emission-line maps presented from here on.
}
\label{Figure:Mrk900_oiiiflux} 
\end{figure}

\begin{figure}
\centering
\includegraphics[angle=0, width=0.9\linewidth]{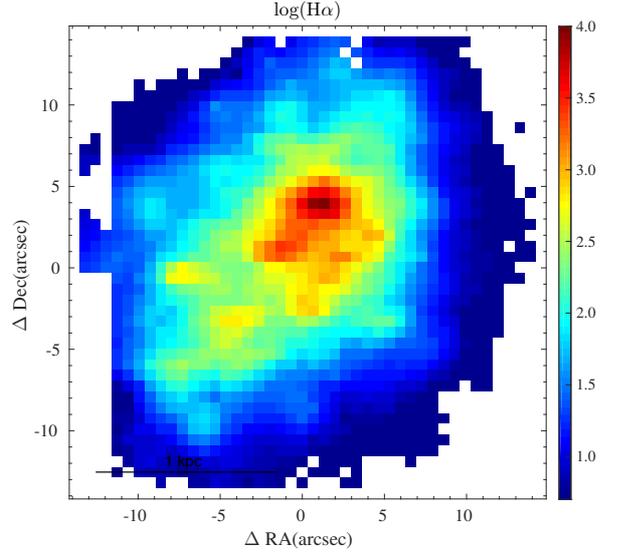}
\caption{H$\alpha$ emission-line flux 
map for Mrk~900.}
\label{Figure:mrk900_haflux} 
\end{figure}

\begin{figure}
\centering
\includegraphics[angle=0, width=0.9\linewidth]{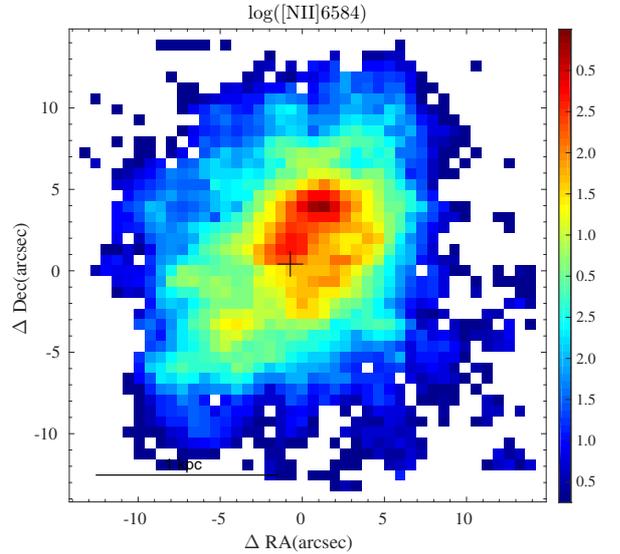}
\caption{[\ion{N}{ii}]~$\lambda6584$ emission-line flux 
map for Mrk~900.}
\label{Figure:Mrk900_niiflux}
\end{figure}

\begin{figure}
\centering
\includegraphics[angle=0, width=0.9\linewidth]{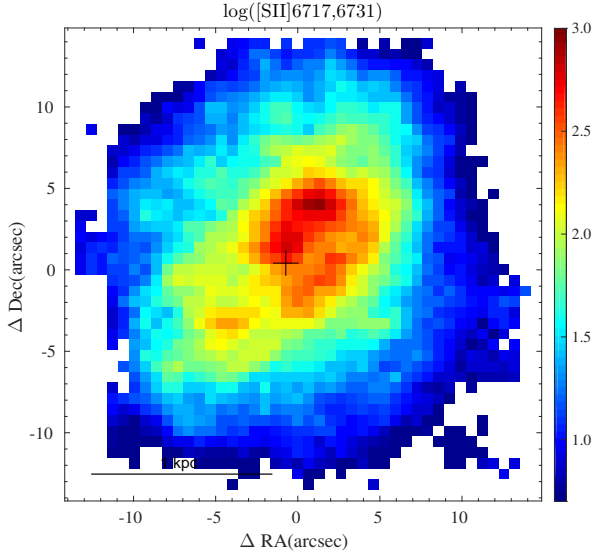}
\caption{[\ion{S}{ii}]~$\lambda\lambda6717,\,6731$ emission-line flux 
map for Mrk~900.}
\label{Figure:Mrk900_siiflux} 
\end{figure}

Emission lines in starburst galaxies trace regions of warm ionized gas, and hence the population of ionizing stars. 
Mrk~900 shows an
irregular morphology in emission lines (see Figures~\ref{Figure:Mrk900_oiiiflux}-\ref{Figure:Mrk900_siiflux}): several SF knots appear  close 
to the galaxy center and along the minor axis (southeast--northwest).  The brightest knot, which spatially   coincides with 
knot~{\sc a} in the broad-band frames,  is displaced about 350~pc northwest from the
continuum peak (at a
distance of
18.9~Mpc the spaxel element translates into 61~pc). 

\smallskip

These SF regions  are embedded in an extended envelope of  diffuse
ionized gas (DIG), which
fills almost the whole FoV. Filaments and curvilinear features are conspicuous
at faint brightness levels:  the larger ones   extend in the SW direction and 
 NE direction. A hole in the warm gas emission, with a diameter
of about 260~pc,  is visible to the NW close to the biggest SF region.

\smallskip

 We adopt here the common definition for the DIG, namely diffuse ionized hydrogen outside the discrete (well-defined) \ion{H}{ii} regions \citep{Reynolds1973, Dopita2003,Oey2007}. To make a clear-cut distinction between spaxels belonging to \ion{H}{ii} regions and to the DIG is not  straightforward (see, e.g.,  the discussion in  \citealt{FloresFajardo2009}). However a substantial fraction of the warm ionized emission in Mrk~900 belong to the DIG 
according to the currently most used criteria: for instance, the 
high values of the low-ionization [\ion{S}{ii}]~$\lambda\lambda6717,\,6731$/\Ha{} line ratio (see Section~\ref{choques} below) and the low values of the  H$\alpha$ equivalent width  (Fig.~\ref{Figure:Mrk900_haeqw}) 
in the outer galaxy regions are characteristic of the DIG following the criteria proposed  in \cite{Blanc2009} and   \cite{Lacerda2018}, respectively. 

\smallskip 

The galaxy displays the same morphological pattern  in all emission
lines, but fainter SF clumps appear better delineated in the high-excitation
[\ion{O}{iii}]~$\lambda5007$ line  than in the hydrogen recombination lines, and they are
only barely visible in the low-ionization  [\ion{N}{ii}]~$\lambda6584$ and
[\ion{S}{ii}]~$\lambda\lambda6717,\,6731$ maps.

\subsection{Interstellar extinction pattern}
\label{extinction}

\begin{figure}
\centering
\includegraphics[angle=0, width=0.9\linewidth]{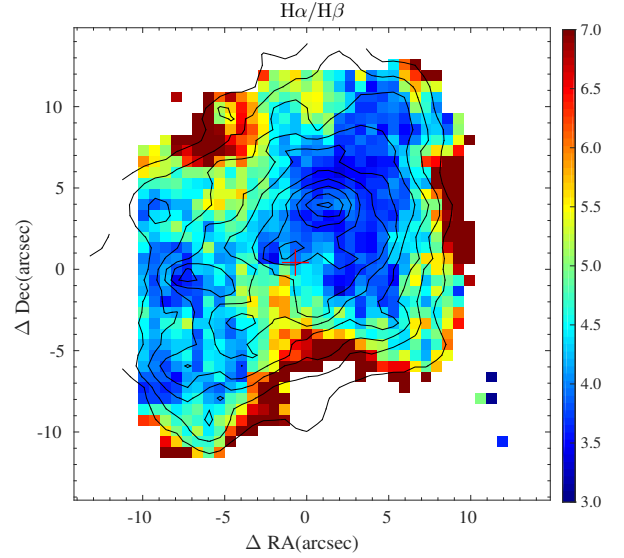}
\caption{H$\alpha$/H$\beta$ ratio map for Mrk~900, with the contours of the H$\alpha$ flux map overplotted. }
\label{Figure:mrk900_reddening} 
\end{figure}

The interstellar extinction in a nebulae is computed by comparing the observed Balmer
decrement values to the theoretical ones \citep{Osterbrock2006}. Since  the Balmer line ratios are
well known from atomic theory,  deviations from the predicted values are assumed to be due
to dust;   the interstellar extinction coefficient, C(H$_{\beta}$), is 
derived as 
\begin{equation}
\frac{F_{\lambda}}{F(H_{\beta})}=\frac{F_{\lambda,0}}{F(H_{\beta,0})}\times10^{-C(H_{\beta})[f(\lambda)-f(H_{\beta})]}
\label{ext_eq}
,\end{equation}
\noindent where 
$F_{\lambda}/F(H_{\beta}$) is the observed ratio of Balmer emission-line intensities 
relative to H$_{\beta}$, $F_{\lambda,0}/F(H_{\beta,0}$)  the 
theoretical ratio and
$f(\lambda$) the adopted extinction law. We applied  Eq.~(\ref{ext_eq})
 to every  spaxel to obtain an
interstellar extinction map (Figure~\ref{Figure:mrk900_reddening}).

\smallskip

In practice, we derived  the extinction coefficient using  only the H$\alpha$/H$\beta$ ratio; although H$\gamma$ also falls 
into the observed wavelength range, it has a poorer S/N  and is also
more severely affected by underlying stellar absorption  compared to H$_{\alpha}$ and H$_{\beta}$
 \citep{Olofsson1995,GonzalezDelgado1999a}. We adopted case~B recombination and 
 low-density limit with a temperature of 10000~K for 
 $F_{\lambda,0}/F(H_{\beta,0}$) in  Eq.~(\ref{ext_eq}). We used the Galactic extinction law from
\cite{ODonnell1994};  Mrk~900 is closer in metallicity to the Large Magellanic Cloud (LMC), but differences between the Galactic and the LMC extinction curves are insignificant in the observed spectral range \citep{Dopita2003}.  

\smallskip 

The interstellar extinction map of Mrk~900
(Figure~\ref{Figure:mrk900_reddening}) reveals a clear 
spatial pattern. 
The lowest extinction values are reached at the position of the SF regions, and the highest  in the
inter-knot area. In addition, extinction anticorrelates with the H$\alpha$
intensity: the maximum extinction values are in the regions of fainter H$\alpha$ emission. These results are consistent with dust being destroyed or swept away  by the most
massive stars.

\smallskip 

In the whole FoV, H$_{\alpha}$/H$_{\beta}\geq$3.5, which indicates a significant  amount of
dust even at the position of the SF regions (H$\alpha$/H$\beta$=3.5 implies an extinction
coefficient C(H$\beta$)=0.23 and a color excess $E(B-V)$=0.17). Away from the SF
regions,  H$_{\alpha}$/H$_{\beta}$ reaches values up to 6, meaning C(H$\beta$)=0.87 and $E(B-V)$=0.64. Such large spatial variability in the extinction agrees with the results of  previous
analyses \citep{Lagos2014,Cairos2015, Cairos2017a,Cairos2017b} and strengthens  the conclusions of
those works: applying a unique extinction coefficient to the whole galaxy, as usually done for
long-slit spectroscopic observations, can lead to large errors in derived fluxes,
magnitudes, and SFR.

\subsection{Excitation and ionization mechanisms}
\label{choques}

Specific emission-line ratios, namely [\ion{O}{iii}]~$\lambda5007$/\Hb{},
[\ion{N}{ii}]~$\lambda6584$/\Ha, [\ion{S}{ii}]~$\lambda\lambda6717,\,6731$/\Ha{}, and 
[\ion{O}{i}]~$\lambda6300$/\Ha{},  reveal the  excitation and ionization mechanisms
 in a nebula \citep{Dopita2003,Osterbrock2006}; IFU observations 
enable us to  investigate  the spatial
variations of excitation and ionization conditions in galaxies
\citep{Sharp2010,Rich2011,Rich2012,Rich2015,Cairos2017a,Cairos2017b,Mingozzi2019}. 

\smallskip

\begin{figure}
\centering
\includegraphics[angle=0, width=0.9\linewidth]{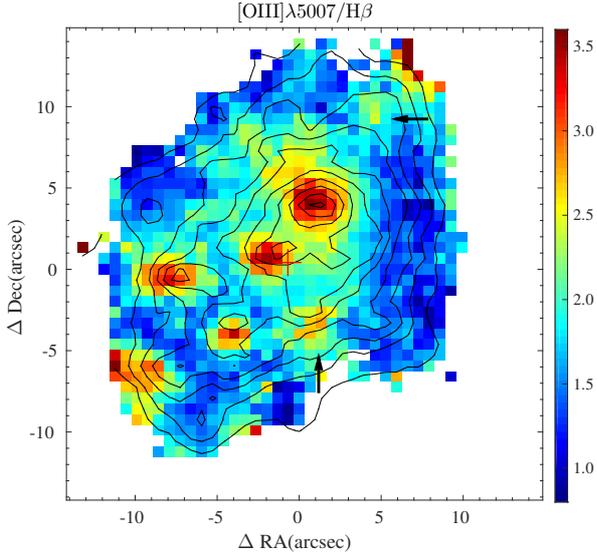}
\caption{[\ion{O}{iii}]~$\lambda5007$/\Hb{} emission-line ratio map with contours on 
H$\alpha$ overplotted. The arrows indicate areas of enhancement of the line ratio, which 
do not spatially coincide with any SF regions.}
\label{Figure:diagnostic-oiii} 
\end{figure}

The [\ion{O}{iii}]~$\lambda5007$/\Hb{} distribution of Mrk~900 (Figure~\ref{Figure:diagnostic-oiii}) is not 
uniform, but  shows a clear spatial pattern: the highest
excitation is reached at the position of the SF regions. This is expected as a harder
ionization source ($h\nu\geq$ 35.1~eV), such as very young O stars, is required to cause a strong
[\ion{O}{iii}]~$\lambda5007$ line. 
The excitation reaches its maximum at the position of the major H$\alpha$
emitter ([\ion{O}{iii}]~$\lambda5007/\Hb{}\sim$3.0). Interestingly, there are  two other
zones, not co-spatial  with any SF knot, with a relatively high
excitation (Figure~\ref{Figure:diagnostic-oiii}). An enhancement on the excitation in regions not
photoionized by stars can be due to the presence of shocks and/or  a large amount of dust; 
the excitation maximum at the south is very close to the peak in interstellar
extinction.

\smallskip 

\begin{figure}
\centering
\includegraphics[angle=0, width=0.9\linewidth]{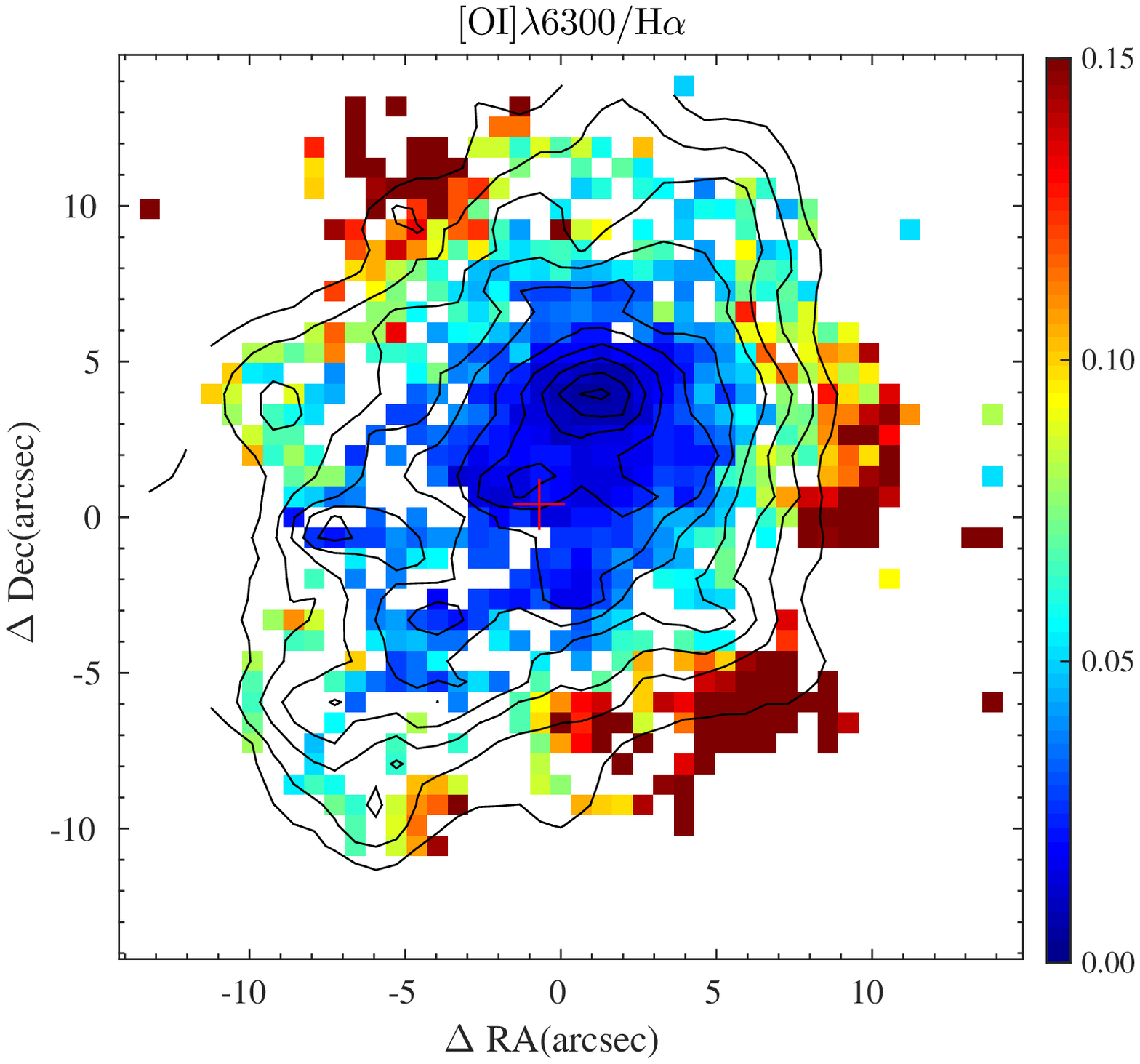}
\caption{[\ion{O}{i}]~$\lambda6300$/\Ha{} emission-line ratio map with contours on 
H$\alpha$ overplotted.}
\label{Figure:diagnostic-oi} 
\end{figure}

\begin{figure}
\centering
\includegraphics[angle=0, width=0.9\linewidth]{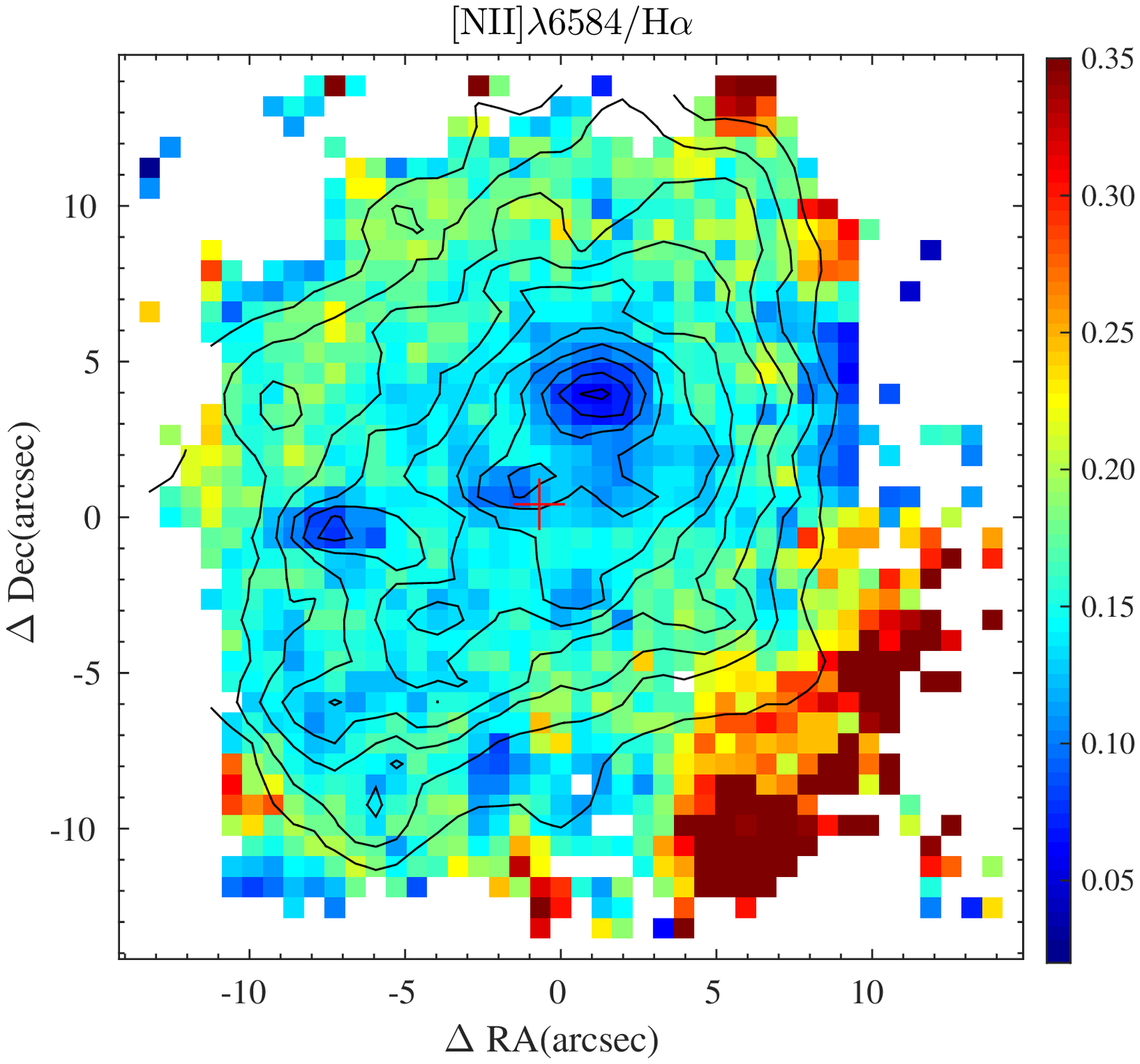}
\caption{[\ion{N}{ii}]~$\lambda6584$/\Ha\ emission-line ratio map with contours on 
H$\alpha$ overplotted.}
\label{Figure:diagnostic-nii} 
\end{figure}

\begin{figure}
\centering
\includegraphics[angle=0, width=0.9\linewidth]{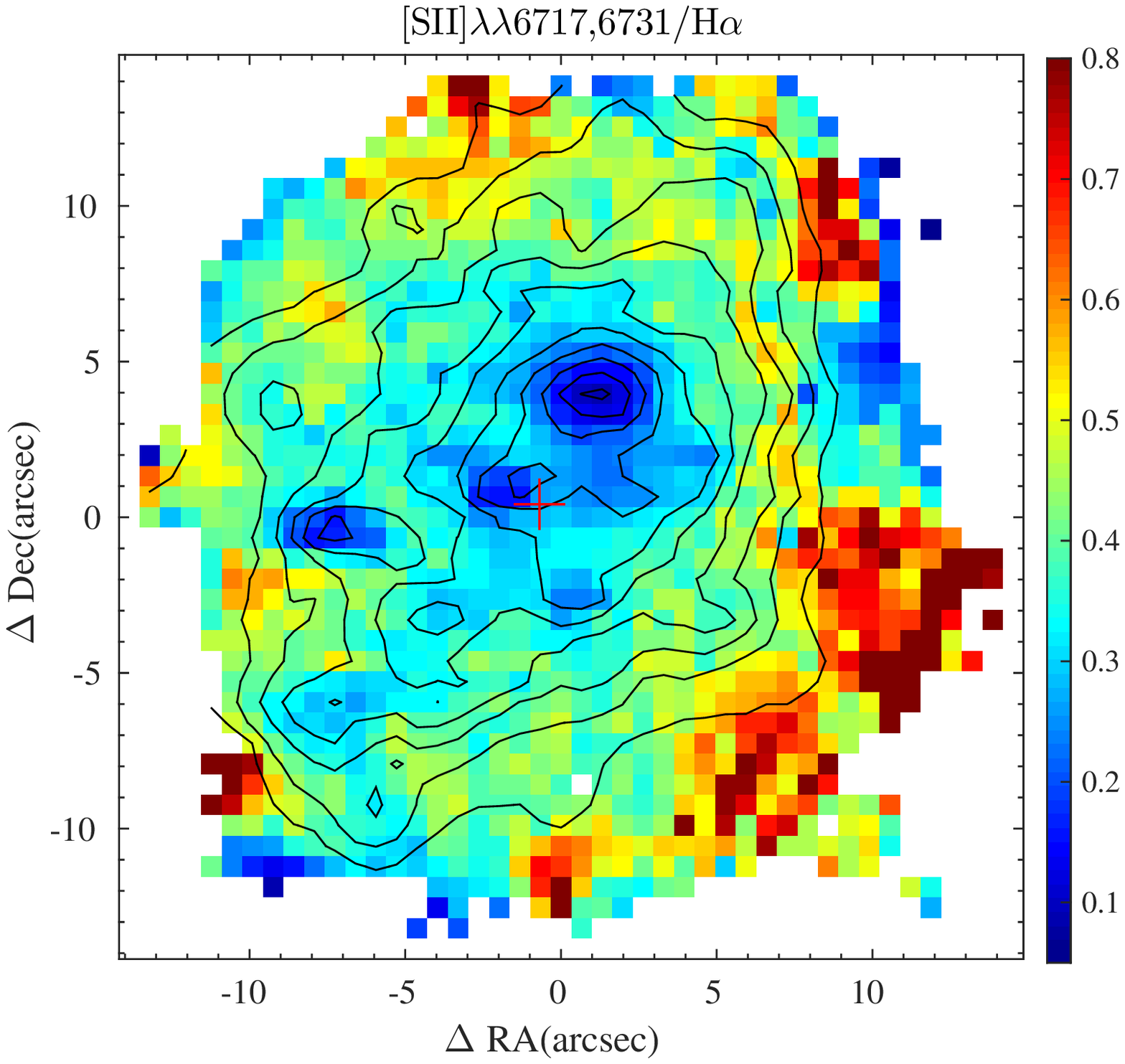}
\caption{[\ion{S}{ii}]~$\lambda\lambda6717,\,6731$/\Ha{} emission-line ratio map with contours on 
H$\alpha$ overplotted.}
\label{Figure:diagnostic-sii} 
\end{figure}

The maps for low-ionization species 
[\ion{O}{i}]~$\lambda6300$/\Ha{}, [\ion{N}{ii}]~$\lambda6584$/\Ha{}, and
[\ion{S}{ii}]~$\lambda\lambda6717,\,6731$/\Ha{} (Figures~\ref{Figure:diagnostic-oi} -- \ref{Figure:diagnostic-sii}, respectively)
show the opposite 
behavior, with the line ratios reaching their minimum at the center of the SF
knots. This is  expected in regions ionized by UV photons coming from massive stars. 
All  three line ratios appreciably
increase outward ([\ion{S}{ii}]~$\lambda\lambda6717,\,6731$/\Ha{} $\geq$ 0.6), 
which are values much higher than those predicted for stellar photoionization.

\smallskip

From the emission-line ratio maps, we  generated spaxel 
resolved diagnostic diagrams;  we plotted in the classical diagnostic
diagrams \citep{Baldwin1981,Veilleux1987}  the value of the line ratios at each individual element of spatial resolution.  This technique permits us to explore the power sources acting in different galaxy regions \citep{Sharp2010,Rich2011,Rich2012,Rich2015,Leslie2014,Cairos2017a,Cairos2017b}.

\smallskip 

\begin{figure*}
\centering
\includegraphics[angle=0, width=0.75\linewidth]{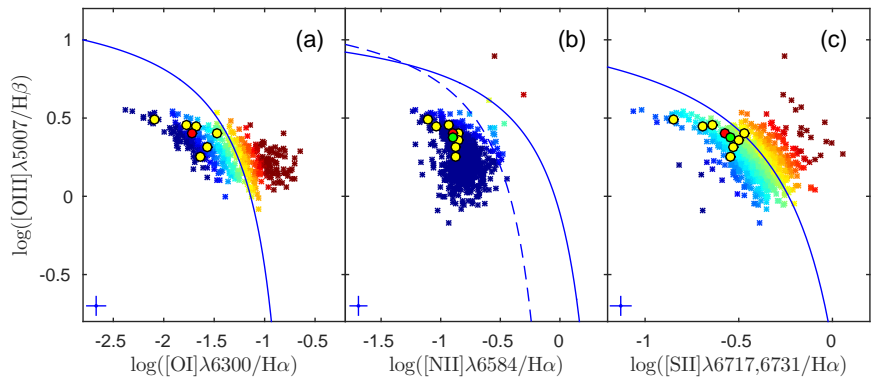}
\caption{Optical emission-line  diagnostic
diagram  for the individual spaxels in Mrk~900;  the  ratio for the SF regions identified 
in the galaxy and for the nuclear and integrated spectrum  are also shown in the plot (yellow, red, and green circles represent the SF regions, the nuclear,  and the integrated spectrum,  respectively). The solid black line in the panels delineates the theoretical ``maximum starburst line''  
derived by \cite{Kewley2001}; the dashed black curve in panel (b) traces the \cite{Kauffmann2003} empirical classification line. To better visualize the results on the diagram, the points are
color-coded according to their distance to the maximum starburst line. In both diagrams, the lower-left section
of the plot is occupied by spaxels in which the dominant energy source is the
radiation from hot stars (blue points in the figure). Additional ionizing mechanisms
shift the spaxels to the top right and right part of the diagrams (from yellow to red). 
}
\label{Figure:diagnostic-spaxel} 
\end{figure*}

Figure~\ref{Figure:diagnostic-spaxel}  displays 
the diagnostic line diagrams for Mrk~900, together with the maximum starburst line (or photoionization line) derived by \cite{Kewley2001};  this line traces the limit between gas photoionized by young stars and gas ionized via other mechanisms (AGN or shock excitation). 
We found  that a considerable fraction of the spaxels fall outside of the area occupied by star photoionization in the diagrams [\ion{O}{iii}]~$\lambda5007$/\Hb{} {versus}  [\ion{O}{i}]~$\lambda6300$/\Ha{} and  [\ion{O}{iii}]~$\lambda5007$/\Hb{} {versus} [\ion{S}{ii}]~$\lambda\lambda6717,\,6731$/\Ha{}. The ratio
[\ion{N}{ii}]~$\lambda6584$/\Ha{}, weakly dependent on the hardness of the radiation but strongly
dependent on the metallicity, is not effective in separating shocks from photoionized regions. In
particular, at low metallicities (0.2~Z$\odot\leq$~Z~$\leq0.4~$Z$\odot$) this diagram is degenerated,
and  shock-ionization and photoionization overlap \citep{Allen2008,Hong2013}. 

\smallskip

\begin{figure}
\centering
\includegraphics[angle=0, width=0.8\linewidth]{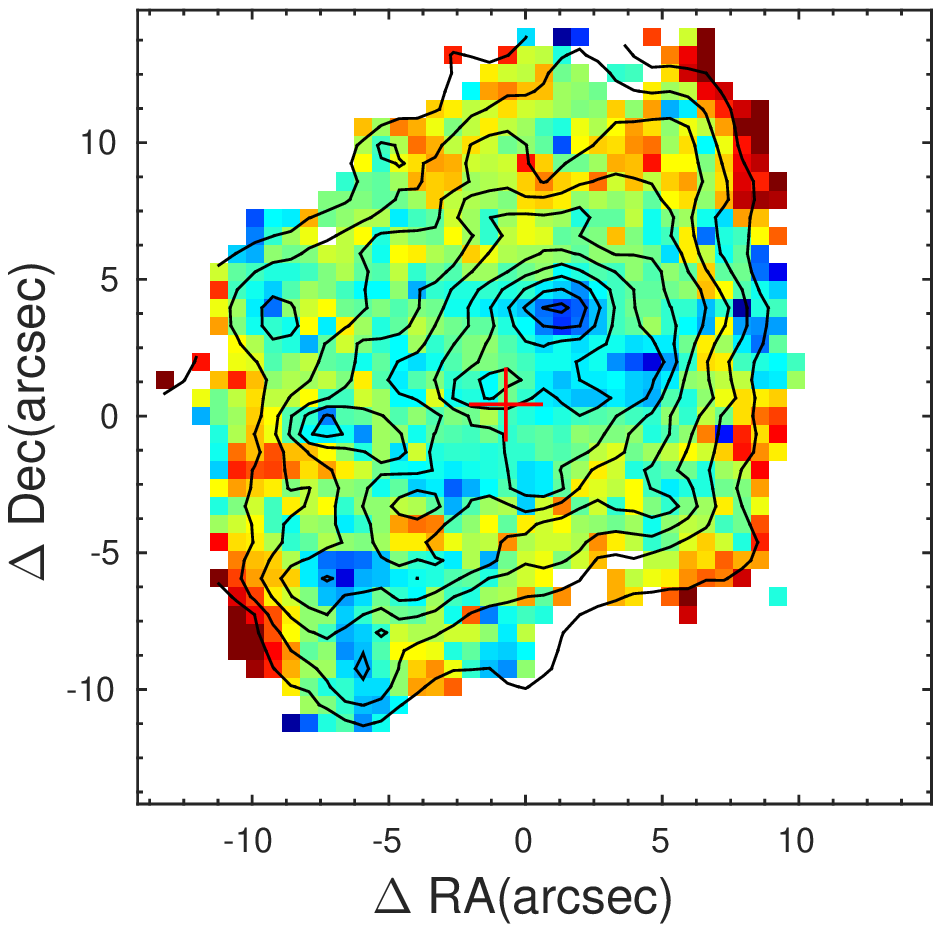}
\caption{Spatial localization of the spaxels in the diagnostic diagram [\ion{O}{iii}]~$\lambda5007$/\Hb{} vs.
[\ion{S}{ii}]~$\lambda\lambda6717,\,6731$/\Ha{}. The color-coding is the same 
as in Figure~\ref{Figure:diagnostic-spaxel}. The redder regions are 
those situated the above the maximum starburst line from \cite{Kewley2001}. }
\label{Figure:map-diagnostic} 
\end{figure}

In order to better visualize this result, we display the spatial position of these spaxels 
on the galaxy in Fig.~\ref{Figure:map-diagnostic}. The areas not photoionized are situated mainly at the periphery of the mapped region and in the inter-knot space, which conforms to the idea that  they are shocked regions, generated in the interface 
between the expanding bubbles (produced by the massive stellar feedback) and the ambient ISM. This behavior has also been observed in other BCGs, for example Haro~14 and Tololo~1937-423 \citep{Cairos2017a, Cairos2017b}, and in dwarf irregular galaxies, for example   NGC4449 \citep{Kumari2017} and   NGC~5253 \citep{Calzetti1999,Calzetti2004}.

\subsection{Ionized gas kinematics}
\label{kinematics}

\begin{figure*}
\centering
\includegraphics[angle=0, width=0.8\linewidth]{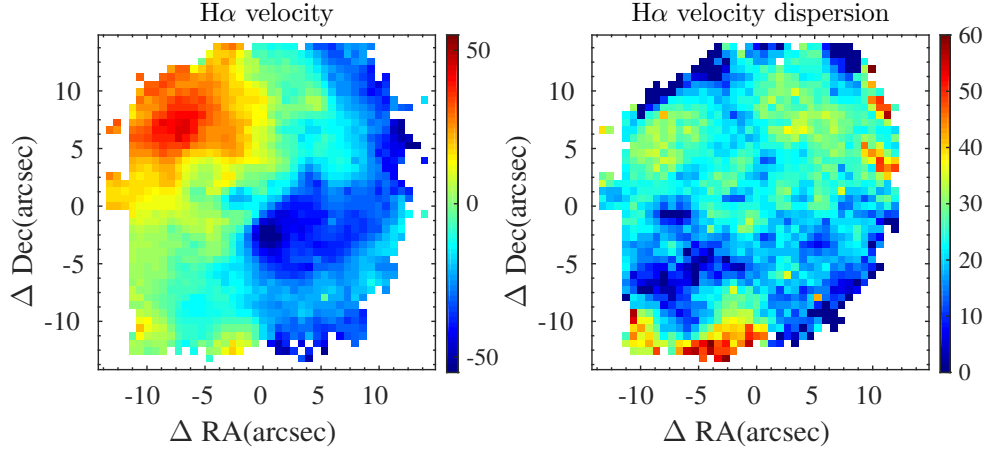}
\caption{H$\alpha$  line-of-sight velocity
field and velocity dispersion map; shown are velocity and velocity dispersions in km~s$^{-1}$.}
\label{Figure:velocitymaps} 
\end{figure*}

The line-of-sight (LOS) velocity map in H$\alpha$, 
measured from the Doppler shift of the line profile centroid
relative to the galaxy systemic velocity, is shown in 
Figure~\ref{Figure:velocitymaps}. A distorted, but 
ordered rotation field pattern is evident, with the velocity gradient 
aligned along the  optical major axis: the regions situated northeast 
are moving away from us, while the southwest regions are moving toward us.  
Deviations from the global rotation are apparent across the whole FoV;  
particularly interesting is the curvilinear feature in the southwest, spatially coincident with the regions of enhancement in [\ion{N}{ii}]~$\lambda6584$/\Ha{} and [\ion{S}{ii}]~$\lambda\lambda6717,\,6731$/\Ha{} in the diagnostic maps. 

\smallskip 

The kinematics of Mrk~900 has been studied in 
the multi-pupil IFS analysis of 18 BCGs by \cite{Petrosian2002}. 
These authors found a homogeneous velocity field in the H$\alpha$ emission line
and excluded the possibility of a strong rotational gradient in the galaxy, but this result clearly arises from their limited FoV. Their velocity map, centered roughly where   
the continuum peaks, 
covers about  9.1$\times$13~arcsec; consequently, the whole field is dominated by the starburst emission.  Using  \ion{H}{i} synthesis observations,  \cite{VanZee2001}
investigated the kinematics of Mrk~900 in a much larger FoV;  the overall pattern of their velocity map  (see their Figure~6) 
is in good agreement with our results.

\smallskip 

The H$\alpha$ LOS velocity dispersion,  derived from the width of the Gaussian fit to the line profile after
accounting for  the instrumental ($\sim$89.9~km/sec) and thermal broadening ($\sim$9.1~km/s), also shows a clear spatial pattern (Figure~\ref{Figure:velocitymaps}). The minimum 
values are reached at the  SF knots 
(5~km~s$^{-1}$ $\leq$ $\sigma$ $\leq$ 15~km~s$^{-1}$), 
while the velocity dispersion
increases up to values 
$\sim$60~km~s$^{-1}$
at the galaxy periphery and in the inter-knot region. 
A nonuniform  spatial distribution  has also been  found 
in several BCGs for which 2D maps of the ionized gas velocity 
dispersion has been published \citep{Bordalo2009,Moiseev2012,Moiseev2015,Cairos2015,Cairos2017a,Cairos2017b}. 
Based on this result,  \cite{Moiseev2015} claimed that the 
large velocity dispersions (turbulent motions in the ionized
gas) observed in dwarf galaxies do not 
reflect virial motions, but are instead related to 
the feedback from young massive stars.

\subsection{Integrated spectroscopy}
\label{integratedspec}

\begin{figure} 
\centering
\includegraphics[angle=0, width=0.9\linewidth]{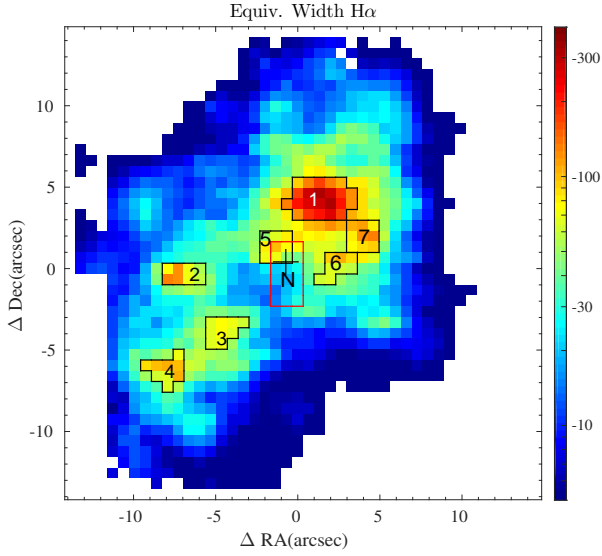}
\caption{H$\alpha$ equivalent width   
map (in \AA) for Mrk~900, with the seven major regions of SF labeled.}
\label{Figure:Mrk900_haeqw} 
\end{figure}

\begin{figure}
\centering
\includegraphics[angle=0, width=1\linewidth]{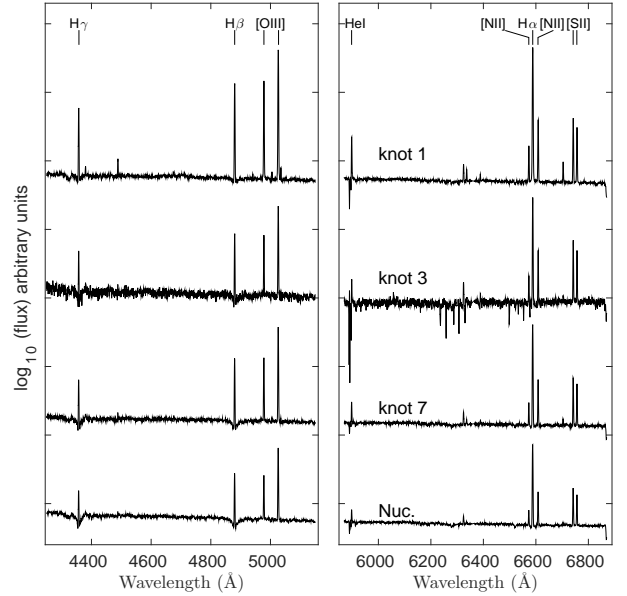}
\caption{Flux-calibrated spectra of three of the seven
 selected \ion{H}{ii} regions and 
the nuclear region in Mrk~900, 
in logarithmic units.}
\label{Figure:spectra} 
\end{figure}

Using the bidimensional maps  we  identified 
the major regions  of SF in the galaxy. To 
delimit their  borders we  used 
the H$\alpha$ equivalent width map; equivalent widths of hydrogen recombination lines are excellent
age indicators  \citep{Dottori1981,Stasinska1996,Leitherer2005,Levesque2013}.  The H$\alpha$ equivalent width map of Mrk~900 (Figure~ \ref{Figure:Mrk900_haeqw}) traces
the H$\alpha$ flux, but the \ion{H}{ii} regions are better delineated and, in particular, various small clumps located at the filaments are conspicuous. 

\smallskip 

Because there is no clear-cut criterion  to set the limits of the knots,
we integrated over a boundary tracing the morphology of
the cluster, taking into account that we are limited by the seeing. In this way we singled out 
seven major \ion{H}{ii} regions in Mrk900 (labeled in Figure~\ref{Figure:Mrk900_haeqw}). We 
generated  the integrated spectrum  of each knot by adding the spectra of the corresponding spaxels, so 
we obtained a higher S/N spectrum compared to those of the individual
fibers,  more suitable to derive physical parameters and abundances. We also produced the ``nuclear'' spectrum, adding
the signal of the spaxels around the continuum peak, and the total spectrum, adding all the spaxels in the H$\alpha$ emission line flux map. 

\smallskip

\begin{figure} 
\centering
\includegraphics[angle=0, width=0.9\linewidth]{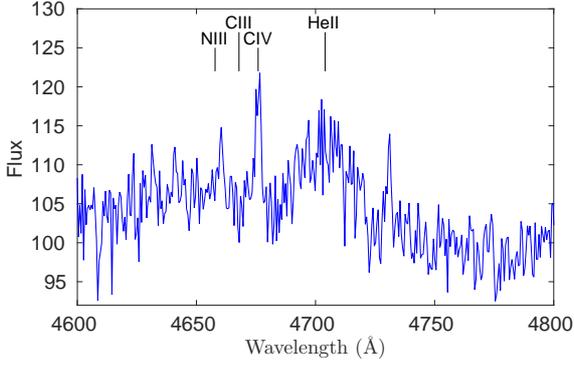}
\caption{Enlarged region of the spectrum around blue WR bump ($\lambda4650-4690$\AA) in the knot~1.  This feature is attributed to the blends of NIII~$\lambda$4640, CIII~$\lambda$4650, CIV~$\lambda$4658, and HeII~$\lambda$4686\AA\ lines.}
\label{Figure:Mrk900_wr} 
\end{figure}

All SF knots display a typical nebular spectrum,
with strong Balmer lines and [OIII], [NII], and [SII] forbidden lines in emission, on top of
an almost featureless continuum. 
As expected, the spectrum of the nuclear region 
presents a higher continuum and more pronounced absorption features.
As an illustration, we show the spectra of three out of the seven SF regions and the 
nuclear region  in Figure~\ref{Figure:spectra}. In knot~1 we  detect the broad-band Wolf-Rayet (WR) bump at 
$\lambda4650-4690$ \AA, the unresolved blend of 
NIII~$\lambda$4640, CIII~$\lambda$4650, CIV~$\lambda$4658, and HeII~$\lambda$4686\AA\ lines (see Fig.~\ref{Figure:Mrk900_wr}; \citealp{Conti1991, Schaerer1999}).
 The S/N  of the spectra is too low for a reliable measurement of the WR bump flux, but just the detection of WR stars strongly constrains the age, duration, and initial mass function (IMF) of
the SF episode
\citep{Meynet1995,Schaerer1998, Guseva2000}. At the metallicity of Mrk~900 (see Section~\ref{abundances} below) the presence of WRs is consistent with these 
stars being formed in an instantaneous burst, with ages between 3 and 6.3~Myr,  and an IMF with an upper mass limit of 100~M$\odot$ \citep{Leitherer1995,Leitherer1999}.  

\subsubsection{Emission-line fluxes}

We measured the observed emission-line fluxes in each of these
spectra using the {\sc iraf} task {\tt splot}. The contribution
of the Balmer lines in absorption was taking into account, as discussed in
Section~\ref{creatingmaps}. From the observed Balmer decrement  we computed the 
interstellar extinction coefficient, as explained in  
Section~\ref{extinction}. 
The reddening-corrected intensity ratios are presented in
Table~\ref{tab:fluxes}, together with the derived interstellar extinction
coefficient, C(H$\beta$); for an easier comparison with other data, the number
of magnitudes of extinction in V, A$_{V}$, and the color excess, $E(B-V)$,  are
also shown.  

\smallskip

\subsubsection{Diagnostic line ratios, physical parameters, and abundances}

\label{abundances}

The values of the diagnostic line ratios,  electron densities ($N_{\rm e}$), and electron temperatures
($T_{\rm e}$)  for the selected regions  and the integrated spectrum of Mrk~900 are shown in 
Table~\ref{tab:diagnostic}. The values of  $N_{\rm e}$ and $T_{\rm e}$ were
computed using the five-level atom {\sc fivel} program in the {\sc iraf nebular} package
\citep{deRobertis1987,ShawDufour1995}. 
The electronic densities were estimated  from  the
[\ion{S}{ii}]~$\lambda\lambda6717,\,6731$ line ratio \citep{Osterbrock2006}; we found 
that all regions present
values in the low-density regime; knot~5 has a slightly higher density, close to 100~cm$^{-3}$. Where the  [\ion{O}{iii}]$\lambda$4363 line could be measured (knots~1,~3,
and 5) we derived the  electron temperature from the
[\ion{O}{iii}]$\lambda$4363/($\lambda$4959+$\lambda$5007) line ratio.

\smallskip 

The oxygen abundance could not be determined using the direct $T_{\rm e}$ method (not even in the three
knots where [\ion{O}{iii}]$\lambda$4363 was measured) because this method  requires the measurement of the
[\ion{O}{ii}]~$\lambda$3727+3729 line, which unfortunately falls out of the VIMOS spectral
range.  Alternatively, the auroral [\ion{O}{ii}]~$\lambda\lambda$7320,7330 lines 
could be used, but for Mrk~900 these features
fall very close to the edge of the VIMOS spectrum in  a zone that is highly affected by sky
residuals, which prevents us from obtaining reliable flux values. We therefore estimated the oxygen
abundances by adopting the empirical method introduced
by \cite{PilyuginGrebel2016}, 
 which  utilizes the intensities of the strong lines
[\ion{O}{iii}]$\lambda\lambda$4957,5007, [\ion{N}{ii}]$\lambda\lambda$6548,6584, and
[\ion{S}{ii}]$\lambda\lambda$6717,6731. This calibrator provides (as usual) separate relations for high-metallicity and low-metallicity objects, but this degeneration can be simply broken using the N$_{2}$ line ratio (N$_{2}$=[\ion{N}{ii}]$\lambda\lambda$6548,6584)/H$\beta$). For Mrk~900, all the measured spectra (i.e., the SF regions and the nuclear and integrated spectra) lie in the upper branch of the calibrator; the division between the upper and lower branches takes place at log N$_{2}$= -0.6. 
The relative accuracy of the abundance 
derived using this method is 0.1~dex. 

\smallskip

The oxygen abundances are similar in all 
the knots and the integrated spectrum, 12+log(O/H)$\sim$8.25  ($\approx$0.3~Z$\odot$\footnote{With solar abundance, 
12+log(O/H)=8.69 \citep{Asplund2009}}); this value  is slightly higher than 
that  found by \cite{Zhao2010} applying the direct-$T_{\rm e}$ method, 12+log(O/H)=8.07$\pm$0.03. Nearly homogeneous chemical abundances have been found for most of the blue compact and dwarf irregular galaxies investigated so far \citep{Croxall2009,Haurberg2013,Lagos2014,Lagos2016,Cairos2017a,Cairos2017b}.

\section{Discussion}

The results derived in the previous section make it possible to
investigate the star-forming history  (SFH) and the SF process in Mrk~900.  The first step used to probe the SFH in a galaxy is to disentangle its stellar populations; it is     
already difficult to work with galaxies that cannot be resolved into stars,  
and this task becomes particularly tricky when we deal with starburst objects. IFS observations, supplying
a broad set of observables for every element of spatial resolution, provide a
novel and powerful way to approach the problem 
\citep{Cairos2017a,Cairos2017b}.

\smallskip

Combing our VIMOS/IFU  and 
broad-band imaging data we distinguish  at least  two  stellar populations in Mrk~900: a  very
young component (exposed in the emission-line maps) and a significantly older
stellar population (traced to large galactocentric distances in the  broad-band pictures). 
We can constrain the properties of these different stellar
populations by comparing the derived observables with the 
predictions of evolutionary synthesis models.

\smallskip

We focus first on the very young stars. Only O and early B stars, with temperatures higher than 30.000~K
and masses above 10~M$_{\odot}$,  produce  photons with energy high enough to
ionize hydrogen. Models show that
these stars  evolve quickly: 
ionizing stars cannot be older than 10~Myr 
\citep{Leitherer1995,Ekstroem2012,Langer2012}. Hence, the star clusters 
generating the \ion{H}{ii} regions in the central parts of Mrk~900 are younger than 10~Myr.

\smallskip 

The H$\alpha$ flux in the 
whole mapped area  (derived from the summed spectrum)  is 5.7$\pm$0.03$\times10^{-13}$erg~s$^{-1}$cm$^{-2}$, consistent with the value 5.5$\pm$0.6$\times10^{-13}$erg~s$^{-1}$cm$^{-2}$ 
reported by \cite{GildePaz2003} from narrow-band images. 
From the total luminosity L(H$\alpha$)=2.4$\times$10$^{40}$~erg~s$^{-1}$, we derive a SFR=0.17$\pm$0.02~M$\odot$~yr$^{-1}$,  applying  the following expression from \cite{Hunter2010}: 
%
$${\rm SFR}(M_{\odot}~{\rm yr}^{-1})=6.9\times10^{-42}\,L_{H\alpha}({\rm erg~s}^{-1}).$$
%
This formula, derived in   \cite{Kennicutt1998}, was modified for a subsolar metallicity, which is the most appropriate for dwarf galaxies. 
The derived SFR is slightly higher than the value 0.12~M$\odot$~yr$^{-1}$ reported  in  \cite{Hunt2015}\footnote{The SFR in \cite{Hunt2015} is derived using H$\alpha$ and 24~$\mu$m following \cite{Calzetti2010}.}.
\smallskip 

 To compare this SFR  with those of other galaxies, 
we must take into account the galaxy size (see discussion in   \citealp{HunterElmegreen2004}). Normalizing  
the  SFR to the radio at surface brightness level 25~mag~arcsec$^{-2}$ ($R_{25}$) 
we obtain SFR=0.0105~M$\odot$~yr$^{-1}$~kpc$^{-2}$, a substantially high SFR even among the BCG class; for a sample of 23~BCGs, 
\cite{HunterElmegreen2004} reported
an average SFR of 0.0062~M$\odot$~yr$^{-1}$~kpc$^{-2}$.

\smallskip 

The global SFR  (0.17$\pm$0.02~M$\odot$~yr$^{-1}$)
and the mass of atomic hydrogen (M$_{HI}$=1.55$\times$10$^{8}$~M$\odot$; \citealp{VanZee2001}) imply a depletion time (the timescale to exhaust the current gas supply of the galaxy;  \citealp{Roberts1963}) $\tau$=0.9~Gyr.   According to the 
classical definition, this classifies Mrk~900  as a  starburst galaxy, i.e., a galaxy whose ``SFR cannot be sustained for a significant fraction of the Hubble time with the available interstellar gas'' \citep{Gallagher2005, Heckman2005,McQuinn2010a}.

\smallskip

\begin{table*}
\caption{H$\alpha$ derived quantities for the major SF regions identified in Mrk~900.}
\begin{center}
\begin{tabular}{|c|c|c|c|c|c|c|c|}
\hline
Knot       &  F(H$\alpha$)                                              & log[L(H$\alpha$)]       &  SFR                             &  W(H$\alpha$) & Age\\
           & (10$^{-16}$~erg cm$^{-2}$s$^{-1}$)      & (erg~s$^{-1}$)            &        M$\odot$ ~yr$^{-1}$     &   (\AA)       & (Myr)  \\\hline\hline
1          &  1602$\pm$37                                                      &    39.83$\pm$0.14       &  0.047$\pm$0.007               & 214              & 5.5 \\
2          &  95$\pm$14                                                 &          38.61$\pm$0.20         &  0.0028$\pm$0.0006            & 94               & 6.3 \\
3          &  95$\pm$11                                                 &          38.61$\pm$0.18         &  0.0028$\pm$0.0006             & 78               & 6.5 \\
4          &  32$\pm$6                                                   &          38.13$\pm$0.23       &  0.0009$\pm$0.0002               &77                & 6.5\\
5          &  382$\pm$186                                                       &          39.21$\pm$0.51        &  0.011$\pm$0.006                 &73                & 6.6 \\
6          &  118$\pm$6                                                 &          38.70$\pm$0.15        &  0.0035$\pm$0.0005                & 78               & 6.5\\
7          &  147$\pm$9                                                 &          38.80$\pm$0.15        &  0.0043$\pm$0.0006               &107               & 6.2\\
\hline
\end{tabular}       
\end{center}
Note: H$\alpha$ fluxes were corrected from interstellar extinction using the values provided in Table~\ref{tab:fluxes}
\label{SF-KnotI}
\end{table*}

\smallskip

The
H$\alpha$ fluxes, luminosities, and equivalent widths of the 
individual \ion{H}{ii} regions identified in Mrk~900 (see Section~\ref{integratedspec})
are shown in Table~\ref{SF-KnotI}.
To constrain their properties, we compare the H$\alpha$ equivalent widths 
with the predictions of 
the {\sc Starburst~99} evolutionary synthesis models \citep{Leitherer1999}. Adopting the models
with metallicity Z=0.008 (the value closest to the metallicity derived
from the emission-line fluxes) we find that we can reproduce the measured equivalent
widths with an instantaneous burst of SF, a Salpeter 
IMF with an upper mass limit of 100 M$\odot$, and ages ranging from 5.5 to 6.6 Myr.
This age range must be understood, however, as an upper limit:  
the 
measured  equivalent widths can decrease as a result of absorption from 
A-F stars 
and/or dilution due to the continuum from an older stellar population \citep{Fernandes2003,Levesque2013}.  
We   corrected the
H$\alpha$ measurements in  Table~\ref{SF-KnotI} for stellar absorption
(see Section~\ref{linefitting}), but no attempt was made to correct for 
the presence of  the older stars; the uncertainties in the ages of the SF knots due to the 
contribution from an older population were estimated by \cite{Cairos2002,Cairos2007}
to be up to $\sim$1-1.5~Myr. 
\smallskip 

The ionized gas emission is manifestly dominated by knot~1: it generates $\sim$30\% of the total H$\alpha$ flux in the observed area. With a  H$\alpha$ luminosity of 6.84$\pm$0.9 $\times$10$^{39}$ergs$^{-1}$ and a
diameter about 300~pc, knot~1 is classified as a giant extragalactic \ion{H}{ii} region
(GEHR). It is, indeed, comparable in size with the 30~Doradus nebula, the
largest GEHR in the Local Group.  The detection of the WR bump in this knot provides an independent age estimation of 3-6 Myr  \citep{Schaerer1998}, consistent with that derived from H$\alpha$.

\smallskip

The morphology of the central starburst region in the continuum  (Figure~\ref{Figure:mrk900_cont}) strongly  differs from the morphology in emission lines (Figures~\ref{Figure:Mrk900_oiiiflux}-\ref{Figure:Mrk900_siiflux}). 
From the seven \ion{H}{ii}  regions identified in emission lines, only knot~1 has a visible counterpart in the continuum (knot~{\sc a}); however, it appears only as a moderate continuum emitter. 
The continuum maximum (knot~{\sc c}), located about the center of the elliptical host, is about 130~pc southwest from its nearest  \ion{H}{ii} region (knot~5). 
Such morphological patterns may indicate the presence of distinct bursts of SF and  spatial migration of the SF sites \citep{Petrosian2002}. By carrying out similar IFU analyses in the BCGs Haro~14 and Tololo~1937-423, we identified two temporally and spatially separated bursts whose ages suggest  a scenario of triggered SF  \citep{Cairos2017a,Cairos2017b}. In 
Mrk~900, however,  the presence of a second episode of SF cannot 
be definitely confirmed, as  
knot~{\sc c} could be also the nucleus of the host component and the increase 
in intensity the result of a higher stellar density. 
The high values of the  H$\gamma$ and H$\beta$ equivalent widths in absorption ($\sim$5\AA) in the nuclear spectrum might indicate the presence of intermediate-age stars,  but the possibility that 
such large equivalent widths originate in the 
nearby  \ion{H}{ii} regions and/or  are due to a strong dilution in the continuum peak cannot be ruled out.

 \medskip 

We now focus on the properties of the galaxy host, 
very well traced in the NOT images. In the galaxy outskirts,  we find no SF regions but a rather  red and regularly shaped host, with elliptical isophotes.   
The outer regions of the SBP are well described by an exponential function; consistently, 
the velocity field indicates the presence of a  rotating disk, although perturbed in the inner parts.
The region covered by VIMOS is still heavily affected by the starburst, and
we could not derive the rotation curve; however,  from the velocity field we estimated a maximum velocity amplitude $\sim$80--90~km~sec$^{-1}$ along the optical major axis. This value is in very good agreement with the velocity 82.0~km~sec$^{-1}$ at a distance 
of $\sim$1~kpc reported by \cite{VanZee2001} from  \ion{H}{i} data. 

\smallskip 

A comparison of the colors of the host galaxy with the predictions of evolutionary synthesis models \citep{Vazdekis1996,
Vazdekis2010,Fioc1997,LeBorgne2004}
suggests ages of several Gyr. 
Thus, the morphology, structure, dynamics, and age of the host galaxy
are similar to those presented by dE galaxies \citep{LinFaber1983,vanZee2004a,vanZee2004b}.

\medskip

In summary, we  identified in Mrk~900 
a very young population ($\leq$6.6~Myr) resolved in an ensemble of \ion{H}{ii} regions and extending about 1~kpc along the galaxy minor axis. This young component presents a rather distorted appearance in emission lines, with multiple filamentary and bubble-like structures. 
Underlying the young stars there is 
an old (several Gyr)  stellar population with smooth elliptical isophotes, which extends up to radius of 4~kpc.

\medskip 

What mechanism has ignited the actual SF burst in Mrk~900 
after (most probably) several 
Gyr of inactivity is not evident.
The nature of the starburst trigger in low-mass systems  
has been (and is still) widely debated  \citep{Pustilniketal2001,Brosch2004}. 
Several 
internal processes have been discussed, for example stochastic self propagating star formation (SSPSF, \citealp{Gerola1980}) or torques in spiral disk clumps \citep{Elmegreen2012}. However, 
an increasing amount of observations suggests  that external triggers,  such as 
interactions and mergers \citep{Brinks1990,Taylor1996,Ostlin2001,Pustilniketal2001,Ekta2008,Lelli2014} or external gas infalling  \citep{LopezSanchez2012,Nidever2013,Ashley2014,Miura2015,Turner2015} play a major role in the ignition of the starburst in  BCGs. These observational results
are strengthened by numerical simulations that  successfully reproduce 
the merger
\citep{Bekki2008} or gas infall scenario \citep{Verbeke2014}.

\smallskip

The observational evidence  speaks against interactions or mergers with a massive galaxy as  being responsible for the present-day SF in Mrk~900: in the  {\sc hyperleda} database the galaxy appears  classified as ``isolated'' (i.e., it does not have bright neighboring objects); the LSB component  shows a very regular behavior,  without tidal features or  any other sign of asymmetry, 
in the optical  and NIR (Section~\ref{photometry} and \citealp{Micheva2013a, Janowiecki2014});  the distortion of the  isophotes  in the central area is clearly due to the superposition of the SF episode; the velocity field is relatively smooth out of the starburst regions, both in the optical (Section~\ref{kinematics}) and in radio \citep{VanZee2001};  finally,  we did not find any significant metallicity variations
among the individual stellar clusters  (Section~\ref{integratedspec}).

\smallskip

On the other hand, interactions with low-mass companions,  mergers between gas-rich dwarf galaxies or external gas infalling are all triggering scenarios consistent with our results: a centrally concentrated SF, as we observe  in Mrk~900,  is expected 
after an interaction/merger event;  a collision can drive mass inflows to the galaxy 
central regions  and ignite a nuclear starburst \citep{Mihos1994, Bekki2008}; 
the alignment of the SF with the galaxy minor axis points   to  
inflowing (generated in a collision) or outflowing gas,  but the two scenarios are rather difficult to distinguish observationally.  The presence of shocked  and  high velocity-dispersion regions (Section~\ref{kinematics}; Section~\ref{choques})  and the substantial amount of dust 
(Fig.~\ref{Figure:mrk900_reddening}) 
is also suggestive of  an interaction or accretion event.

\smallskip 

In addition, the \ion{H}{i} synthesis observations of Mrk~900  presented 
in \cite{VanZee2001} reveal an extended and distorted morphology. The galaxy neutral gas distribution  
shows clear deviations from symmetry in the outer regions, the most evident being two extensions
departing toward the east  and northeast (Fig. 6 in \citealp{VanZee2001}).
More recent work, dealing with higher sensitivity and spatial resolution observations on 
nearby starbursting dwarfs, have resolved such distorted  \ion{H}{i} morphologies  into different components, for example   tails, plumes, or clouds
(IC~10, \citealp{Ashley2013, Nidever2013}; NGC~5253, \citealp{Kobulnicky2008,LopezSanchez2012} or NGC~1569, \citealp{Johnson2012,Johnson2013}), and  interpreted these findings as evidence of interactions or inflowing gas. In the case of  BCG NGC~5253, the infall scenario is  further supported by kinematic CO observations \citep{Miura2015,Turner2015}.

\section{Summary and conclusions}

This work presents results on a spectrophotometric analysis of the BCG Mrk~900,
carried out  by combining  VIMOS/IFU observations with deep (BVR) broad-band imaging. From the IFU data  we built
continuum, emission-line, and  diagnostic line ratio maps, and generated LOS velocity and velocity dispersion maps of the central starburst region. 
Using  the broad-band frames, which trace the underlying stellar host, we derived SBPs and color maps. 

\smallskip 

From our analysis we highlight the following results:

\medskip 

$\bullet$ We  disentangled two stellar components in Mrk~900: an  ionizing (very young) population, resolved in individual stellar clusters and extending about 1~kpc along the galaxy minor axis, and a very regular LSB stellar host, which reaches galactocentric distances up to $\sim$4~kpc and exhibits red colors, consistent with ages of several Gyr.

\medskip 

$\bullet$  
We  generated the integrated spectrum of the major seven \ion{H}{ii} regions  identified in Mrk~900,  and of its nuclear region. From these spectra 
we derived reliable physical parameters and oxygen abundances. We found, for all knots,  similar values of the abundance, 12+log(O/H)$\sim$8.25  ($\approx$0.3~Z$\odot$), and no evidence of metallicity variations. Using evolutionary synthesis models we estimated  ages of  5.5-6.6~Myr for all the ionizing clusters. 
We detected in the larger \ion{H}{ii} region (knot~1) the WR bump at 
$\lambda4650-4690$ \AA,\ and therefore we demonstrated that  Mrk~900  can be classified as a  WR galaxy. 

\smallskip 
$\bullet$ We showed that Mrk~900 contains a substantial amount of dust, with A$_{V}\geq$0.48 in the whole mapped area. The dust distribution is  inhomogeneous, with a dust lane crossing the central starburst northeast--southwest; dust lanes and patches are also distinguished in the broad-band frames and color map.

\smallskip

$\bullet$ Diagnostic maps and diagnostic diagrams have shown the presence of  shock-dominated  zones in Mrk~900; these regions are situated primarily at the periphery of the mapped (starburst) area and in the inter-knot region,  conforming with the idea of shocks being generated in the interface 
between expanding bubbles (generated by the massive stars) and the ambient ISM.

\smallskip

$\bullet$ We  built velocity and velocity dispersion fields from the brightest emission lines. Although deviation from circular motions is evident, the galaxy displays an overall rotation pattern. The dispersion map is inhomogeneous, and the areas of higher dispersion coincide spatially with the areas of low surface brightness. 

\smallskip

$\bullet$ Given our observational results, we argue that an interaction with a low-mass system, a merger between gas-rich dwarf galaxies, or infalling from external gas clouds are all plausible scenarios for the ignition of the actual burst of SF in Mrk~900.

\begin{landscape}
\begin{table}
\footnotesize
\caption{Reddening-corrected line intensity ratios, normalized to H$\beta$, for the SF knots and the nuclear region in Mrk~900.\label{tab:fluxes}}
\begin{tabular}{lccccccccc}
\hline
Ion                                        &\multicolumn{1}{c}{knot~1}    &\multicolumn{1}{c}{Knot~2}     &\multicolumn{1}{c}{Knot~3}      &\multicolumn{1}{c}{Knot~4}     &\multicolumn{1}{c}{Knot~5}    &\multicolumn{1}{c}{Knot~6}        &\multicolumn{1}{c}{Knot~7}      & \multicolumn{1}{c}{Nuc}    & \multicolumn{1}{c}{Sum}\\
\hline                                                                                                                                                                                                                                                               
4340~H$\gamma$                      &  0.482$\pm$0.004                      &  0.434$\pm$0.020                         &  0.589$\pm$0.029                         &  0.498$\pm$0.042                       &  0.494$\pm$0.009                         &  0.474$\pm$0.016                           &  0.485$\pm$0.019                               &  0.498$\pm$0.011              &  0.493$\pm$0.003\\
4363~[OIII]                                 &  0.021$\pm$0.002                      &  ---                                                     &  0.037$\pm$0.013                  &  ---                                                 &  0.045$\pm$0.012                            &   ---                                                    &  ---                                                            &    ---                                          &  ---            \\
4472~HeI                                    &  0.038$\pm$0.002                       &  ---                                                    &    ---                                            &  ---                                                   &  0.034$\pm$0.008                         &   ---                                                    &  ---                                                     &   ---                                          & ---     \\
4861~H$\beta$                               &  1.000                                        &  1.000                                           &  1.000                                            &  1.000                                                 &  1.000                                                   &    1.000                                             & 1.000                                                       &   1.000                                  & 1.000\\                                     
4959~[OIII]                                 &  1.043$\pm$0.006                      &  0.930$\pm$0.030                         &  0.834$\pm$0.028                          &  0.751$\pm$0.039                      &  0.958$\pm$0.014                     &  0.608$\pm$0.013                          &  0.696$\pm$0.016                         &  0.825$\pm$0.012               & 0.791$\pm$0.003      \\
5007~[OIII]                                 &  3.091$\pm$0.016                       &  2.771$\pm$0.074                        &  2.534$\pm$0.062                  &  2.288$\pm$0.088                     &  2.864$\pm$0.031                             &  1.794$\pm$0.030                          &  2.047$\pm$0.034                                &  2.509$\pm$0.030              &2.401$\pm$0.007     \\
5875~HeI                                    &  0.102$\pm$0.002                       & 0.098$\pm$0.021                         &   0.088$\pm$0.013                          &  0.093$\pm$0.032                     &  0.088$\pm$0.006                             &  ---                                                    &  0.085$\pm$0.007                                 &   0.091$\pm$0.007             & --- \\
6300~[OI]                                   &  0.024$\pm$0.001                       &  0.070$\pm$0.014                        &  0.097$\pm$0.009                          &  ---                                                 &  0.050$\pm$0.005                            &  0.066$\pm$0.008                          &  0.077$\pm$0.007                               &  0.056$\pm$0.006              & ---     \\
6548~[NII]                                 &  0.074$\pm$0.001                        &  0.088$\pm$0.011                        &  0.130$\pm$0.012                          &  0.137$\pm$0.037                      &  0.103$\pm$0.004                        &  0.124$\pm$0.007                               &  0.126$\pm$0.007                                 &    0.118$\pm$0.006           & 0.135$\pm$0.002  \\
6563~H$\alpha$                     &  2.870$\pm$0.020                       &  2.870$\pm$0.104                        &  2.870$\pm$0.095                  &  2.870$\pm$0.146                       &  2.870$\pm$0.043                        &  2.870$\pm$0.060                            &  2.870$\pm$0.061                          &   2.870$\pm$0.047             & 2.870$\pm$0.012  \\
6584~[NII]                                 &  0.226$\pm$0.002                       &  0.270$\pm$0.015                        &  0.403$\pm$0.017                            &  0.404$\pm$0.032                     &  0.337$\pm$0.008                        &  0.381$\pm$0.012                                 &  0.385$\pm$0.011                             &   0.367$\pm$0.008             & 0.361$\pm$0.002    \\
6678~HeI                                  &  0.030$\pm$0.001                        &  ---                                                  &  ---                                                 &  ---                                                    &  0.022$\pm$0.005                        &  ---                                                         &  0.025$\pm$0.007                                 &     ---                                        & ---   \\
6717~[SII]                                &  0.238$\pm$0.003                        &  0.354$\pm$0.018                        &  0.585$\pm$0.024                  &  0.541$\pm$0.040                      &  0.375$\pm$0.008                         &  0.487$\pm$0.014                               &  0.496$\pm$0.013                              &    0.438$\pm$0.009            & 0.477$\pm$0.002    \\
6731~[SII]                                &  0.170$\pm$0.002                        &  0.241$\pm$0.014                        &  0.386$\pm$0.017                  &  0.369$\pm$0.032                      &  0.282$\pm$0.006                          &  0.336$\pm$0.011                               &  0.354$\pm$0.011                               &  0.324$\pm$0.008             & 0.339$\pm$0.002 \\
7136~[ArIII]                               &  0.097$\pm$0.002                      &  0.083$\pm$0.012                         &  0.090$\pm$0.011                  & 0.097$\pm$0.023                        &  0.096$\pm$0.004                         &  ---                                                            &  0.086$\pm$0.008                               &      0.091$\pm$0.006          & ---\\
\hline
F$_{H\beta}$                             &  558$\pm$13                                       &  30$\pm$4                                   &  33$\pm$4                                   &  11$\pm$2                                       &  133$\pm$65                                 &  41$\pm$2                                        &  51$\pm$3                                               & 16$\pm$1                      & 1997$\pm$24\\
C$_{H\beta}$                              &  0.439$\pm$0.006                         &  0.467$\pm$0.030                          &  0.434$\pm$0.028                       &  0.398$\pm$0.042                        &  0.439$\pm$0.012                       &  0.288$\pm$0.017                                &  0.317$\pm$0.018                                & 0.418$\pm$0.014               & 0.408$\pm$0.004         \\
W(H$\gamma$)$_{ab}$        &  2.1                                                    &  --                                                &      3.4                                               &     ---                                               & 3.7                                                 &   ---                                                     &      1.9                                             &       5.0                                & 2.5             \\
W(H$\beta$)$_{ab}$                 &  1.9                                                   &  2.2                                              &      3.3                                        &     ---                                               & 4.5                                                  &   2.4                                                       &       3.7                                                       &        5.1                                 & 3.5          \\
$A_{V}$                                            &  0.945$\pm$0.012                        &  1.012$\pm$0.065                      &   0.938$\pm$0.058                          &  0.860$\pm$0.092                        &  0.948$\pm$0.027                      &  0.623$\pm$0.038                            &  0.685$\pm$0.038                                 &     0.904$\pm$0.030          & 0.882$\pm$0.007 \\                                                                                                                                                                                            
E(B-V)                                            &  0.306$\pm$0.004                        &  0.326$\pm$0.021                       &   0.302$\pm$0.019                         &  0.278$\pm$0.029                        &  0.306$\pm$0.009                       &  0.201$\pm$0.012                           &  0.221$\pm$0.012                           &     0.292$\pm$0.009          & 0.284$\pm$0.002\\
\hline 
\end{tabular}
\end{table}
Notes. Reddening-corrected line fluxes normalized to F(H$\beta$)=1.
The reddening-corrected H$\beta$
flux (in units of 10$^{-16}$erg~s$^{-1}$~cm$^{-2}$), 
the
interstellar extinction coefficient C$_{H\beta}$, 
and the values of the equivalent width in absorption for H$\gamma$ and H$\beta$ 
(in \AA) are also provided in the table. A$_{V}$=2.16$\times$C(H$\beta$) and
$E(B-V)$=0.697$\times$C(H$\beta$) \citep{Dopita2003}.
\end{landscape}

\begin{landscape}
\begin{table}
\footnotesize
\caption{Line ratios, physical parameters, and abundances.\label{tab:diagnostic}}
\begin{tabular}{lccccccccc}
\hline
Parameter                                               &\multicolumn{1}{c}{Knot~1}       &\multicolumn{1}{c}{Knot 2}      &\multicolumn{1}{c}{Knot 3}        &\multicolumn{1}{c}{Knot 4}     &\multicolumn{1}{c}{Knot 5}         &\multicolumn{1}{c}{Knot 6}         &\multicolumn{1}{c}{Knot 7}      &  \multicolumn{1}{c}{Nuclear} &  \multicolumn{1}{c}{Integrated}\\
\hline
$[\ion{O}{iii}]~\lambda5007/\Hb$                         &    3.091$\pm$0.094             &  2.784$\pm$0.573               &   2.534$\pm$0.371                &   2.280$\pm$0.469             & 2.864$\pm$0.190                  &  1.794$\pm$0.115                  & 2.047$\pm$0.142                &  2.508$\pm$0.175               & 2.401$\pm$0.040     \\
$[\ion{O}{i}]~\lambda6300/\Ha$                           &    0.008$\pm$0.001             &  0.021$\pm$0.004               &   0.034$\pm$0.004                &   ---                         & 0.017$\pm$0.002                  &  0.023$\pm$0.003                  & 0.027$\pm$0.003                &  0.019$\pm$0.002               & ---     \\
$[\ion{N}{ii}]~\lambda6584/\Ha$                          &    0.079$\pm$0.001             &  0.092$\pm$0.010               &   0.140$\pm$0.011                &   0.141$\pm$0.017             & 0.118$\pm$0.004                  &  0.133$\pm$0.006                  & 0.134$\pm$0.005                &  0.128$\pm$0.005               &0.126$\pm$0.001       \\
$[\ion{S}{ii}]~\lambda\lambda6717\,6731/\Ha$             &    0.142$\pm$0.002             &  0.202$\pm$0.169               &   0.338$\pm$0.021                &   0.317$\pm$0.031             & 0.229$\pm$0.007                  &  0.287$\pm$0.010                  & 0.296$\pm$0.010                &  0.266$\pm$0.008               & 0.285$\pm$0.002      \\\hline 
N$_{e}$   (cm$^{-3}$)                                    &   $<$100                       &  $<$100                        &  $<$100                          &  $<$100                       &  $\approx$100                        &    $<$100                         &   $<$100                       &    $<$100                    & $<$100 \\
T$_{e}$  (K)                                             &   10236               &  ---                           &    13450                         &   ---                       &   13840                           &         ---                       &    ----                        &         ---                  & ---   \\
12+log(O/H)$^{2}$                                        &    8.23                        &  8.21                          &  8.24                            &   8.26                        & 8.25                              &  8.25                             &  8.25                          &    8.25                      & 8.25 \\
\hline\hline 
\end{tabular}
\end{table}
Notes.  Abundances are derived following 
 \cite{PilyuginGrebel2016}.
\end{landscape}

\begin{acknowledgements} L.M. Cair{\'os} acknowledges support from the Deutsche Forschungsgemeinschaft
(CA~1243/1-1 and CA 1243/1-2). 
We thank  Rafael Manso Sainz  for extremely stimulating discussions and a careful reading of the manuscript.  The data
presented here were obtained [in part] with ALFOSC, which is provided by the Instituto de Astrof\'isica
de Andaluc\'ia (IAA) under a joint agreement with the University of Copenhagen and NOTSA.
This research has made use of the NASA/IPAC Extragalactic
Database (NED), which is operated by the Jet Propulsion Laboratory, Caltech,
under contract with the National Aeronautics and Space Administration. We acknowledge the usage of the HyperLeda database (http://leda.univ-lyon1.fr).
\end{acknowledgements}

\bibliographystyle{aa}
\bibliography{vimos}

\end{document}